\documentclass[twocolumn]{aastex631}
\usepackage{amsmath} 
\usepackage{soul} 
\usepackage{multirow} 
\usepackage{bm} 
\usepackage{amsmath} 
\usepackage{xspace} 
\usepackage{gensymb} 
\usepackage{float}
\usepackage{customcommands} 
\usepackage{newunicodechar,graphicx}
\usepackage{multirow, bigdelim, makecell, booktabs}
\PassOptionsToPackage{hyphens}{url}\usepackage{hyperref}

\DeclareRobustCommand{\okina}{%
  \raisebox{\dimexpr\fontcharht\font`A-\height}{%
    \scalebox{0.8}{`}%
  }%
}
\newunicodechar{ʻ}{\okina}

\begin{document}
\title{HD 119130 b is not an ``ultra-dense'' sub-Neptune}
\correspondingauthor{Joseph M. Akana Murphy}
\email{joseph.murphy@ucsc.edu}
\author[0000-0001-8898-8284]{Joseph M. Akana Murphy}
\altaffiliation{NSF Graduate Research Fellow}
\affiliation{Department of Astronomy and Astrophysics, University of California, Santa Cruz, CA 95064, USA}

\author[0000-0002-4671-2957]{Rafael Luque}
\altaffiliation{NHFP Sagan Fellow}
\affiliation{Department of Astronomy and Astrophysics, University of Chicago, Chicago, IL 60637, USA}

\author[0000-0002-7030-9519]{Natalie M. Batalha}
\affiliation{Department of Astronomy and Astrophysics, University of California, Santa Cruz, CA 95064, USA}

\author[0000-0002-0810-3747]{Ilaria Carleo}
\affiliation{Instituto de Astrofísica de Canarias (IAC), 38205 La Laguna, Tenerife, Spain}
\affiliation{Departamento de Astrofísica, Universidad de La Laguna (ULL), 38206 La Laguna, Tenerife, Spain}

\author[0000-0003-0987-1593]{Enric Palle}
\affiliation{Instituto de Astrofísica de Canarias (IAC), 38205 La Laguna, Tenerife, Spain}
\affiliation{Departamento de Astrofísica, Universidad de La Laguna (ULL), 38206 La Laguna, Tenerife, Spain}

\author[0000-0003-2404-2427]{Madison Brady}
\affiliation{Department of Astronomy \& Astrophysics, University of Chicago, Chicago, IL 60637, USA}

\author[0000-0003-3504-5316]{Benjamin Fulton}
\affiliation{NASA Exoplanet Science Institute/Caltech-IPAC, Pasadena, CA 91125, USA}

\author[0000-0002-9305-5101]{Luke B. Handley}
\affiliation{Department of Astronomy, California Institute of Technology, Pasadena, CA 91125, USA}

\author[0000-0002-0531-1073]{Howard Isaacson}
\affiliation{501 Campbell Hall, University of California at Berkeley, Berkeley, CA 94720, USA}

\author[0000-0002-4197-7374]{Gaia Lacedelli}
\affiliation{Instituto de Astrofísica de Canarias (IAC), 38205 La Laguna, Tenerife, Spain}

\author[0000-0001-9087-1245]{Felipe Murgas}
\affiliation{Instituto de Astrofísica de Canarias (IAC), 38205 La Laguna, Tenerife, Spain}
\affiliation{Departamento de Astrofísica, Universidad de La Laguna (ULL), 38206 La Laguna, Tenerife, Spain}

\author[0000-0002-7031-7754]{Grzegorz Nowak}
\affiliation{Institute of Astronomy, Faculty of Physics, Astronomy and Informatics, Nicolaus Copernicus University, Grudzi\c{a}dzka 5, 87-100 Toru\'{n}, Poland}
\affiliation{Instituto de Astrofísica de Canarias (IAC), 38205 La Laguna, Tenerife, Spain}
\affiliation{Departamento de Astrofísica, Universidad de La Laguna (ULL), 38206 La Laguna, Tenerife, Spain}

\author[0000-0003-2066-8959]{J.~Orell-Miquel}
\affiliation{Instituto de Astrofísica de Canarias (IAC), 38205 La Laguna, Tenerife, Spain}
\affiliation{Departamento de Astrofísica, Universidad de La Laguna (ULL), 38206 La Laguna, Tenerife, Spain}

\author[0000-0002-4143-4767]{Hannah L. M. Osborne}
\affiliation{Mullard Space Science Laboratory, University College London, Surrey, UK}
\affiliation{European Southern Observatory, Karl-Schwarzschild-Straße 2, Garching bei München, Germany}

\author[0000-0001-5542-8870]{Vincent Van Eylen}
\affiliation{Mullard Space Science Laboratory, University College London, Surrey, UK}

\author[0000-0001-5664-2852]{María Rosa Zapatero Osorio}
\affiliation{Centro de Astrobiología (CSIC-INTA), Crta. Ajalvir km 4, E-28850 Torrejón de Ardoz, Madrid, Spain}
\begin{abstract}
We present a revised mass measurement for \sysI b (aka K2-292 b), a transiting planet ($P = 17$ days, \rplanet $=$ \rpIb \rearth) orbiting a chromospherically inactive G dwarf, previously thought to be one of the densest sub-Neptunes known. Our follow-up Doppler observations with HARPS, HARPS-N, and HIRES reveal that \sysI b is, in fact, nearly one-third as massive as originally suggested by its initial confirmation paper. Our revised analysis finds \mplanet $=$ \mpIb \mearth (\mplanet $<$ \mpIbupperlim \mearth at 98\% confidence) compared to the previously reported \mplanet $=$ \mpIbLXIX \mearth. While the true cause of the original mass measurement's inaccuracy remains uncertain, we present the plausible explanation that the planet's radial velocity (RV) semi-amplitude was inflated due to constructive interference with a second, untreated sinusoidal signal in the data (possibly rotational modulation from the star). \sysI b illustrates the complexities of interpreting the RV orbits of small transiting planets. While RV mass measurements of such planets may be precise, they are not necessarily guaranteed to be accurate. This system serves as a cautionary tale as observers and theorists alike look to the exoplanet mass-radius diagram for insights into the physics of small planet formation.
\end{abstract}

\section{Introduction} \label{sec:intro}
\citet[][hereafter \luqueXIX]{luque19} reported the discovery and confirmation of the sub-Neptune-size planet \sysI b (aka K2-292 b). \luqueXIX detected the planet's transits in \ktwo photometry \citep{howell14} and subsequently conducted RV follow-up with the Calar Alto high-Resolution search for M dwarfs with Exoearths with Near-infrared and optical \'Echelle Spectrographs \citep[CARMENES;][]{carmenes14, carmenes18}. \sysI b orbits its G dwarf host every 17 days. Jointly modeling the four \ktwo transits and their 18 CARMENES RVs, the authors found \rplanet $=$ \rpIb \rearth and \mplanet $=$ \mpIbLXIX \mearth, making this planet unusually massive for its size. At the time of its publication, \sysI b was one of only two\footnote{According to the data from the TEPCat catalog \citep{southworth11} as shown in Figure 6 from \luqueXIX.} planets with \mplanet $\geq20$ \mearth and \rplanet $\leq3$ \rearth \citep[the other being K2-66 b, a planet in the photoevaporation desert orbiting a subgiant star;][]{sinukoff17}. 

Given its curiously high bulk density of \rhoplanet $=$ \rhoIbLXIX \gcc, \luqueXIX dubbed \sysI b a rare ``ultra-dense'' sub-Neptune. The authors posited that the planet must have formed with high density since it is too weakly irradiated ($S_\mathrm{p} = 67$ $S_\mathrm{\oplus}$) to have lost a massive primordial envelope via photoevaporation \citep[e.g.,][]{owen13}. Furthermore, the inner disk of the planet's G dwarf host would not have contained sufficient material to form a 20 \mearth planet in situ at 0.1 au \citep{schlichting14}, meaning \sysI b must have formed beyond the snow line and migrated inward over time---possibly in lockstep with the snow line itself \citep{kennedy08}. \luqueXIX suggested that the migration could have been caused by Kozai-Lidov oscillations \citep{dawsonChiang14, mustill17}, which may have been triggered by a potential long-period companion associated with their detection of a linear RV trend. In this context, ultra-dense sub-Neptunes may represent a valuable tracer of protoplanetary disk composition beyond the snow line, and their orbital architectures could lend insights into the physics of planet migration.

Since the confirmation of \sysI b, other ultra-dense or ``superdense'' sub-Neptunes have appeared in the literature, including K2-182 b \citep{murphy21} and HD 21749 b \citep{dragomir19, gan21}. To date, there are only eight planets (not including \sysI b) with \mplanet $\geq 20$ \mearth and \rplanet $\leq3$ \rearth, whose masses were measured using RVs.\footnote{According to the NASA Exoplanet Archive's Planetary Systems Table, as accessed on 2024 Sept 02 \citep{nea}, for planets with better than 33\% and 15\% fractional measurement precision in mass and radius, respectively.} There are another three such planets with masses derived from transit timing variations \citep[TTVs; e.g.,][]{haddenLithwick14}.

As with any collection of seemingly unusual planets---for example, planets lying in the photoevaporation desert \citep[e.g.,][]{szabo11, lundkvist16, armstrong20} or the so-called ``super-puffs'' \citep[e.g.,][]{lee16, jontofhutter19, libbyroberts20}---it is natural to wonder if these objects truly represent a unique outcome of planet formation, or if the measurements (of the planet masses, in particular) are the product of some conspiracy between observational biases and model misspecification. \cite{lange24} find that the majority of superdense sub-Neptunes with mass measurements from RV monitoring were confirmed using less than 30 observations. Superdense sub-Neptunes may be particularly vulnerable to publication bias because, for a given measurement uncertainty, a larger mass will lead to a higher-precision detection. This bias may, in turn, serve to influence our understanding of the small planet mass-radius relation \citep{burt18}.

To verify the accuracy of \sysI b's mass measurement, we conducted follow-up Doppler observations as part of the THIRSTEE\footnote{Tracking Hydrates In Refined Sub-neptunes to Tackle their Emergence and Evolution.} survey \citep{lacedelli24arxiv}. In total, we obtained 57 new RVs of \sysI. Combined with the CARMENES data from \luqueXIX, our observations reveal that \sysI b is not, in fact, an ultra-dense sub-Neptune. Instead, we find \mplanet $=$ \mpIb \mearth (\mplanet $<$ \mpIbupperlim \mearth at 98\% confidence), which is much more typical for a planet of its size. We also find that the data prefer including a low-frequency signal, which can either be modeled as an outer planetary-mass companion with $P \approx 200$ days and $K = $ \KIc \mps, or as a long-term linear trend. We explore possible explanations for the original mass measurement's inaccuracy, including untreated stellar activity and imperfect time-sampling.

The paper is organized as follows. In \S\ref{sec:obs}, we describe our observations. In \S\ref{sec:analysis}, we present our RV analysis and revised mass measurement for \sysI b. In \S\ref{sec:explanations}, we explore possible explanations for the inaccuracy of the \luqueXIX mass measurement. In \ref{sec:discussion}, we discuss the planet's possible composition and formation and evolution scenarios in the context of its revised mass estimate. We conclude in \S\ref{sec:conclusion}.

\section{Observations} \label{sec:obs}
Here we summarize the RV observations of \sysI. The data, including the archival CARMENES RVs from \luqueXIX, can be found in Table \ref{tab:rvs}.

\subsection{CARMENES}
\luqueXIX obtained 18 RVs from CARMENES between UT 2018 Jun 14 and 2018 Jul 18. The CARMENES RVs have a median internal uncertainty of 3.1 \mps and an RMS of 7.1 \mps about the mean value.

\subsection{HARPS}
The High-Accuracy Radial velocity Planetary Searcher \citep[HARPS;][]{mayor03} is an ultra-precise optical \'Echelle spectrograph with sub-$\mathrm{m\,s^{-1}}$ precision installed at the European Southern Observatory (ESO) 3.6\,m telescope at La Silla Observatory in Chile. We obtained 31 high-resolution spectra between UT 2023 Apr 02 and 2023 Jul 09 under program 0111.C-0138 (PI: E. Pall\'e). Besides RV measurements, we extracted the full-width at half maximum (FWHM) and bisector span (BIS) of the cross-correlation function (CCF) from the FITS headers, as computed by the DRS ESO HARPS pipeline. The RVs have a median internal uncertainty of 2.3 \mps and an RMS of 5.6 \mps about the mean value.

\subsection{HARPS-N}
The High-Accuracy Radial velocity Planetary Searcher-North \citep[HARPS-N;][]{harpsn} spectrograph is mounted at the 3.6\,m Telescopio Nazionale Galileo of Roque de los Muchachos observatory in La Palma, Spain. We obtained 16 high-resolution spectra between UT 2023 Apr 13 and 2024 May 13 under the observing programs CAT23A\_52 and CAT23B\_74 (PI: I. Carleo). RVs and additional spectral indicators were derived from the CCF using an online version of the DRS pipeline \citep{cosentino14}, the YABI tool\footnote{Available at \url{http://ia2-harps.oats.inaf.it:8000}.}. The RVs have a median internal precision of 1.9 \mps and an RMS of 4.3 \mps about the mean value.

\subsection{\keckhires}
The High Resolution Echelle Spectrometer \citep[HIRES;][]{vogt94} is mounted on the 10\,m Keck I telescope at the W. M. Keck Observatory on Maunakea, Hawai\okina i. We obtained 10 high-resolution spectra of \sysI between UT 2023 Apr 10 and 2023 Jun 25 under observing program 2023A\_U156 (PI: N. Batalha). For these observations, a warm (50\degree\ C) cell of molecular iodine was placed at the entrance slit \citep{butler96}. The superposition of the iodine absorption lines on the stellar spectrum provides both a fiducial wavelength solution and a precise, observation-specific characterization of the instrument's point spread function (PSF). To measure precise RVs using the iodine method, a high-resolution, high-\snr, iodine-free ``template'' spectrum must also be obtained to create a deconvolved stellar spectral template (DSST) of the host star. We obtained the iodine-free template of \sysI on UT 2023 Jun 24. The data collection and reduction followed the methods of the California Planet Search (CPS) as described in \cite{howard10}. The RVs have a median internal precision of 1.5 \mps and an RMS of 2.5 \mps about the mean value.

\begin{deluxetable}{cccccc} \label{tab:rvs}
\tabletypesize{\normalsize}
\tablecaption{\sysI RVs}
\tablehead{
  \colhead{Time} & 
  \colhead{RV} & 
  \colhead{RV Unc.} &
  \colhead{Inst.} \\
  \colhead{[BJD]} & 
  \colhead{[\mps]} & 
  \colhead{[\mps]} &
  \colhead{}
}
\startdata
2458284.416 & 0.86 & 2.14 & CARMENES \\
\nodata & \nodata & \nodata & \nodata \\
\enddata
\tablecomments{The complete set of 75 RVs (i.e., the 18 observations from \citealt{luque19} plus the 57 new observations published in this work) used in our analysis of \sysI. Only the first row of the table is shown here to inform its contents and format. BJD is reported using the TDB standard \citep[e.g.,][]{eastman10}. Model-specific instrumental offsets have not been applied to the RV values listed here. The RV errors listed here represent measurement uncertainty and have not been added in quadrature with the corresponding instrument jitter values resulting from our models of the data. This table is available in its entirety online in machine-readable format.}
\end{deluxetable}

\section{Analysis} \label{sec:analysis}

\subsection{Reproducing the \luqueXIX results} \label{sec:replicate_lxix}
First, we attempted to reproduce the results from \luqueXIX by restricting ourselves to the 18 CARMENES RVs and replicating their RV analysis. \luqueXIX find $K = 6.1 \pm 1.1$ \mps and $e = 0.04^{+0.06}_{-0.03}$ for \sysI b as part of a joint model with the \ktwo photometry. We only analyzed the CARMENES RVs (i.e., we did not recreate a joint model that included the \ktwo photometry) but placed informed Gaussian priors on the orbital period ($P$) and time of inferior conjunction (\transitTime) of \sysI b using the posterior estimates from \luqueXIX. All other model parameters had wide, uninformed priors. To guard ourselves against potential systematic biases between analysis tools, we fit the data using two separate modeling codes, \radvel \citep{radvel} and \exoplanet \citep{exoplanet:exoplanet}. Mimicking the \luqueXIX RV model, we find that $K_\mathrm{b}$ is consistent to within $1\sigma$ of the \luqueXIX value for both codes. 

\subsection{Model comparison} \label{sec:model_comparison}
After reproducing the results from \luqueXIX, we turned our attention to analyzing the entire RV time series from all four instruments. We compared different models of the data using the small sample corrected Akaike Information Criterion \citep[AICc;][]{akaike74, burnham02}. The AIcc is defined as 
\begin{equation} \label{eqn:aicc}
    \mathrm{AICc} = \mathrm{AIC} + \frac{2 N_\mathrm{par} (N_\mathrm{par} + 1)}{N_\mathrm{obs} - N_\mathrm{par} - 1},
\end{equation}
and the Akaike Information Criterion \citep[AIC;][]{akaike74} is given as
\begin{equation} \label{eqn:aic}
    \mathrm{AIC} = 2 N_\mathrm{par} - 2 \ln \mathcal{\hat{L}}.
\end{equation}
$N_\mathrm{par}$ is the number of free model parameters, $N_\mathrm{obs}$ is the number of observations, and $\mathcal{\hat{L}}$ is the model's likelihood function maximized with respect to the model parameters. Throughout, we assume a log-likelihood of the form 
\begin{equation} \label{eqn:likelihood}
    \ln \mathcal{L} = -\frac{1}{2} \sum_{i = 1}^{N_\mathrm{obs}} \Big[\ln (2 \pi \sigma_{i,\:\mathrm{tel}}^2) + \frac{(x_i - \mu_{i,\:\mathrm{tel}})^2}{\sigma_{i,\:\mathrm{tel}}^2}\Big],
\end{equation}
where $x_i$ is the RV of the $i$-th observation taken at time $t_i$. We define
\begin{equation} \label{eqn:jitter}
    \sigma_{i,\:\mathrm{tel}}^2 = \sigma_i^2 + \sigma_\mathrm{jit,\:tel}^2,
\end{equation}
where $\sigma_i$ is the measurement uncertainty on the $i$-th observation and $\sigma_\mathrm{jit,\:tel}$ is the corresponding RV jitter parameter for the instrument that produced the observation. The model's predicted RV at time $t_i$ is given as
\begin{equation} \label{eqn:mean}
    \mu_{i,\:\mathrm{tel}} = \mathrm{Keplerian}(t_i,\: P,\:T_\mathrm{c},\:e,\:\omega_\mathrm{p},\:K) + \gamma_\mathrm{tel}
\end{equation}
where Keplerian(...) is the sum of the RV contributions from all of the Keplerian orbits considered by the model and $\gamma_\mathrm{tel}$ is the corresponding instrument-specific RV offset.

While there are no hard and fast rules that dictate model selection with the AIC (or AICc), \cite{burnham04} provide the following guidelines for comparing the AIC of a reference model, AIC$_0$, and of some other model, AIC$_i$: 
\begin{itemize}
    \item If $\mathrm{AIC}_i - \mathrm{AIC}_0 \equiv \Delta \mathrm{AIC}_i < 2$, the two models are nearly indistinguishable.
    \item If $2 < \Delta \mathrm{AIC}_i < 10$, the $i$th model is disfavored.
    \item If $\Delta \mathrm{AIC}_i > 10$, the $i$th model is ruled out.
\end{itemize}

The AICc is essentially the same as the AIC, but with an additional penalty term for model complexity, since the AIC has a tendency to favor models that overfit when $N_\mathrm{obs}$ is small. \cite{burnham02} recommend using the AICc in place of the AIC when $N_\mathrm{obs} \lesssim 40 \times N_\mathrm{par}$, which is typically the case for models of RVs, where $N_\mathrm{par}$ is usually on the order of 10 and $N_\mathrm{obs} \lesssim 100$. 

\subsection{Selecting a baseline model} \label{sec:baseline}
To model the orbit of \sysI b, we placed an informed Gaussian prior on $P$ using the posterior estimate from \luqueXIX. Assuming a linear ephemeris and propagating the \luqueXIX value for \transitTime to near the median of the HARPS, HARPS-N, and HIRES RV observations, \transitTime $= 2460235.4 \pm 0.1$ BJD. Given the long gap in time between the \ktwo transits and our RV follow-up, we elected to place a uniform prior on \transitTime the with a width of $\pm8$ days, centered at the propagated \luqueXIX value. We found that models performed better when using this wider prior on \transitTime, as opposed to a Gaussian corresponding to the formally propagated uncertainty. The AICc did not prefer an eccentric orbit for \sysI b, and an eccentric orbit did not affect the planet's measured mass, so we enforced a circular orbit. A maximum a posteriori (MAP) fit of the data results in $K_\mathrm{b} = 2.20$ \mps.

\begin{figure}
    \centering
    \includegraphics[width=\columnwidth]{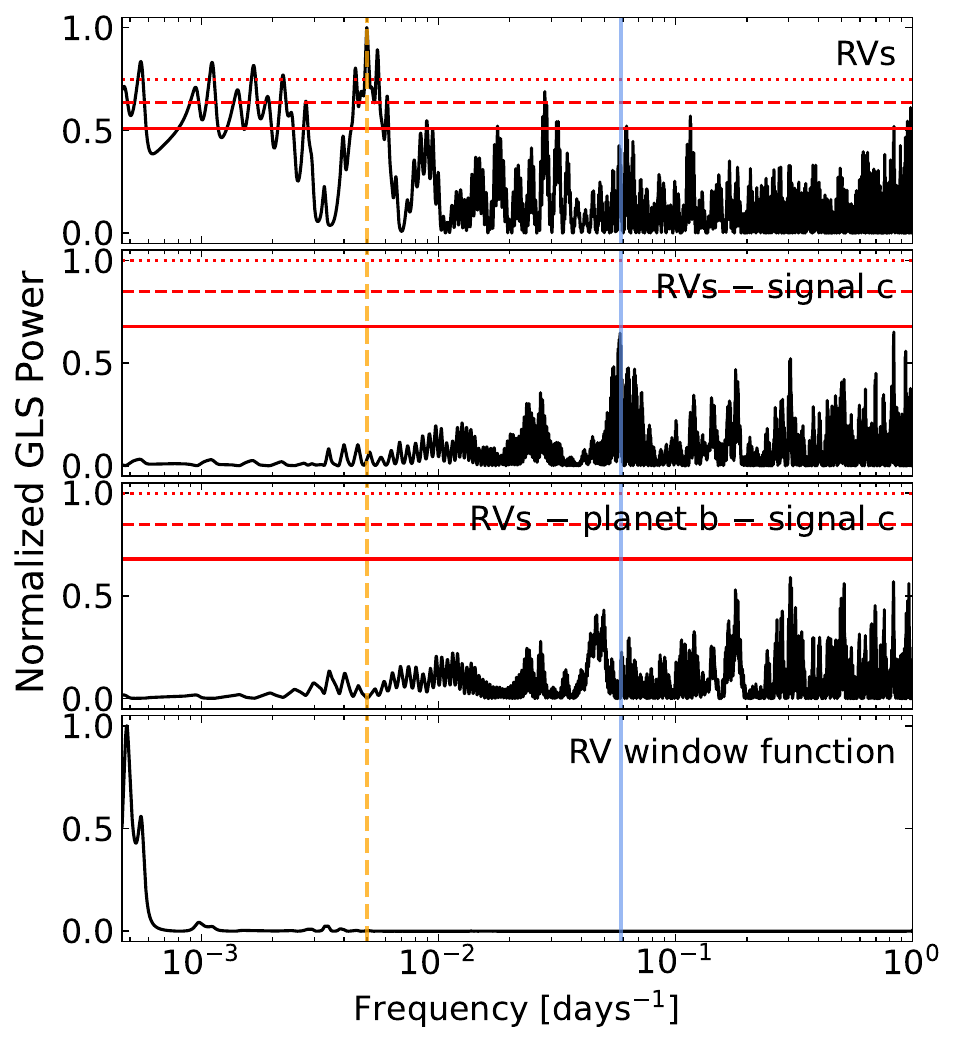}
    \caption{GLS periodograms of the \sysI RVs (minus instrumental offsets), RV residuals, and RV window function. The solid (dashed) blue (orange) vertical line corresponds to the period of planet b (signal c). The dotted, dashed, and solid red horizontal lines correspond to the 0.1, 1, and 10\% FAP thresholds, respectively, for each periodogram, as computed following \cite{baluev08}.}
    \label{fig:hd119130_periodograms}
\end{figure}

After subtracting the orbit of planet b from the data, we computed a Generalized Lomb-Scargle \citep[GLS;][]{zechmeister09} periodogram of the residuals. In the GLS periodogram, we find an envelope of peaks, centered at $P \approx 200$ days, rising above the 0.1\% false-alarm probability \citep[FAP;][]{baluev08} threshold. Fitting a second circular Keplerian orbit to the data, we find a candidate signal with $P = 200.3$ days and $K = 4.9$ \mps. A GLS periodogram of the data after removing this second signal does not produce any peaks that meet the 0.1\% FAP level. GLS periodograms of the data and residuals are shown in Figure \ref{fig:hd119130_periodograms}. The AICc did not prefer an eccentric orbit for the $P = 200$ days signal, and including an eccentric orbit for the outer signal did not affect the mass of planet b. 

What is the nature of this second signal? In both the CARMENES and the HARPS RVs, the data contain a linear trend of similar slope that is inconsistent with the orbit of planet b. \luqueXIX account for the trend in the CARMENES data by adding a linear drift term ($\dot{\gamma}$) to their RV model, and suggest that it may indicate the presence of an outer companion. With our larger data set spanning a much longer baseline, this slope is subsumed by the Keplerian orbit with $P \approx 200$ days. For completeness, we tested a one-planet model that included a linear trend. This configuration did not affect the measured RV semi-amplitude of planet b, but it was ruled out compared to a two-Keplerian fit, with $\Delta$AICc $> 30$.

Comparing a one- and two-Keplerian model using the AICc, we find that a two-Keplerian model is preferred at the $\Delta$AICc $ = 24$ level. We do not claim that this second signal is planetary---additional data are required to reveal its true nature---but it is clear that a single-planet solution does not adequately describe the data. Moving forward, we consider this circular, two-Keplerian configuration as our ``baseline'' model. The model parameters and priors are described in Table \ref{tab:hd119130_model}.
\begin{deluxetable}{lc}
\tablecaption{The \sysI RV model \label{tab:hd119130_model}}
\tabletypesize{\normalsize}
\tablehead{\colhead{Parameter [units]} & \colhead{Prior}}
\startdata
\multicolumn{2}{l}{\emph{Planet b}} \\
$P$ [days] & $\mathcal{N}$(16.9841, 0.0008) \\
\transitTime [BJD] & $\mathcal{U}$[$T_{0,\:\mathrm{b}} - 8$, $T_{0,\:\mathrm{b}} + 8$] \\
$e$ & $\equiv 0$ \\ 
$\omega_\mathrm{p}$ [rad] & \nodata \\ 
$K$ [m s$^{-1}$] & $\mathcal{U}$[0, 100] \\
\\
\multicolumn{2}{l}{\emph{Signal c}} \\
$P$ [days] & $\mathcal{U}$[$P_{0,\:\mathrm{c}} - 20$, $P_{0,\:\mathrm{c}} + 20$] \\
\transitTime [BJD] & $\mathcal{U}$[$T_{0,\:\mathrm{c}} - 100$, $T_{0,\:\mathrm{c}} + 100$]  \\
$e$ & $\equiv 0$ \\ 
$\omega_\mathrm{p}$ [rad] & \nodata \\ 
$K$ [m s$^{-1}$] & $\mathcal{U}$[0, 100] \\
\\
\multicolumn{2}{l}{\emph{Instrument parameters}} \\
$\gamma_\mathrm{tel}$ [m s$^{-1}$] & $\mathcal{U}$[$-250$, $250$] \\
$\ln \sigma_\mathrm{jit,\:tel}$ [$\ln$ m s$^{-1}$] & $\mathcal{U}$[$-3$, $6$]  \\
\enddata
\tablecomments{$\mathcal{N}$(X, Y) refers to a Gaussian distribution with mean X and standard deviation Y. $\mathcal{U}$[X, Y] refers to a uniform distribution on the interval X and Y. The uniform prior on the time of inferior conjunction for planet b is centered on $T_{0,\:\mathrm{b}} = 2460235.4$ BJD, which comes from propagating the \luqueXIX transit time to near the middle of the observing baseline for the RVs collected in 2023 and 2024, assuming a linear ephemeris. The uniform prior on the orbital period of candidate signal c is centered on $P_{0,\:\mathrm{c}} = 200$ days, which represents the largest peak in a periodogram of the one-planet residuals. The center of the prior on the time of inferior conjunction of candidate signal c, $T_{0,\:\mathrm{c}}$, is arbitrarily chosen to be the median of the RV time series. There are offset and RV jitter parameters ($\gamma_\mathrm{tel}$ and $\sigma_\mathrm{jit,\:tel}$, respectively) for each instrument. We fit the natural logarithm of $\sigma_\mathrm{jit,\:tel}$ to avoid pathological posterior geometries known to cause issues for Hamiltonian Monte Carlo-based samplers \citep{neal03, betancourt15}.
}
\end{deluxetable}

\subsection{A systematic search for additional signals} \label{sec:rvsearch_hd119130}
After identifying a baseline RV model for \sysI, we then used \rvsearch \citep{rosenthal21} to systematically search the RV residuals for additional periodicity. \rvsearch compared our baseline, circular, two-Keplerian model to a series of three-Keplerian models, defined by a grid of orbital periods for the third Keplerian. The period grid bounds were 2 days and $5 \times \tau$, with $\tau$ being the observational baseline, in days, of the RV data (in our case, $\tau = 2160$ days). The grid was sampled such that in frequency-space, adjacent grid points were spaced by $(2 \pi \tau)^{-1}$, corresponding to the minimum peak width found in a Lomb-Scargle periodogram \citep{lomb76, scargle82} of unevenly sampled time series data \citep{horne86}. 

\rvsearch computed the difference in the Bayesian information criterion \citep[BIC;][]{schwarz78} between the two- and three-Keplerian models. The BIC is defined as 
\begin{equation} \label{eqn:bic}
    \mathrm{BIC} = N_\mathrm{par} \ln N_\mathrm{obs} - 2 \ln \mathcal{\hat{L}},
\end{equation}
where notation is the same as before. Using the detection methodology described by \cite{howard16}, \rvsearch then estimated the $\Delta$BIC value corresponding to an FAP threshold of 0.1\%. In our case, we find that $\Delta$BIC $= 52.1$ corresponds to FAP $= 0.1\%$. Since none of the three-Keplerian models resulted in a $\Delta$BIC value that exceeded the 0.1\% FAP threshold, we adopted the two-Keplerian solution moving forward.

\begin{figure}
    \centering
    \includegraphics[width=\columnwidth]{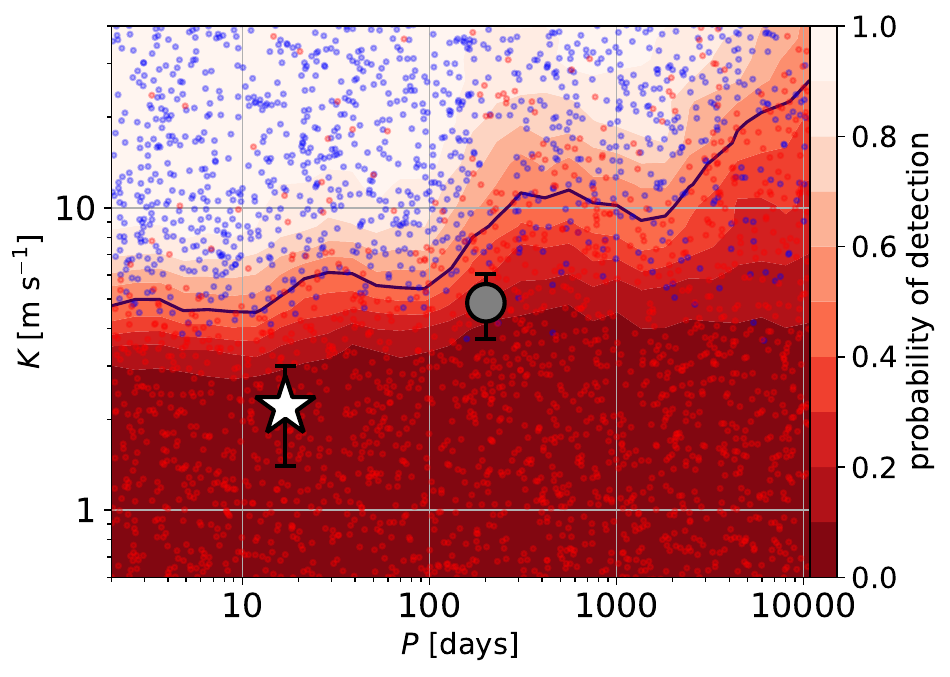}
    \caption{Sensitivity map for our injection-recovery tests on the \sysI RV data using \rvsearch. Points represent individual $P$ and $K$ combinations that were injected into the RV residuals. Blue points correspond to signals that were successfully recovered after removing the orbit of \sysI b and the candidate signal at $P \approx 200$ days, while red points correspond to signals that were not recovered. Signals were considered ``recovered" if their corresponding peak in a periodogram of the residuals rose above the 0.1\% FAP threshold, as determined following the methods of \cite{howard16}. The results of the individual injections and recovery searches were used to compute the contours, which correspond to detection probability (see colorbar). The solid black line corresponds to a detection efficiency of 50\%. The white star represents \sysI b and the gray circle represents the outer candidate, signal c.}
    \label{fig:hd119130_rvsearch_map}
\end{figure}

To get a sense of the range of signals that could be present in the data but fall below our detection threshold, we used \rvsearch to conduct a series of injection-recovery tests (Figure \ref{fig:hd119130_rvsearch_map}). To run an injection-recovery test, we randomly sampled (uniformly in log-log space) an orbital period and an RV semi-amplitude value from a grid spanning $P \in [2, 5 \times \tau]$ days and $K \in [0.5, 50]$ \mps. We then added the corresponding signal to the RVs assuming a circular Keplerian orbit and computed the $\Delta$BIC between the two- and three-Keplerian models of the augmented data. If the $\Delta$BIC between the models exceeded the FAP $= 0.1\%$ threshold, we considered the additional signal to be successfully recovered. We repeated this procedure for 3,000 pairs of $P$ and $K$. 

For $P < 70$ days (i.e., the baseline of the \ktwo observations), we find that the data are not sensitive ($\lesssim 50\%$ recovery rate) to Keplerians with $K \lesssim 5$ \mps. \luqueXIX conducted a box least squares \citep[BLS;][]{kovacs02} search of the \ktwo photometry, but found no transiting planet candidates other than \sysI b, meaning that any additional planets in the system with $P < 70$ days are likely non-transiting. For a planet with $P = 35$ days on a circular orbit, $b > 1$ implies $i_\mathrm{p} < 88.6\degree$, which is consistent with the inclination of \sysI b ($i_\mathrm{p} = 88.4 \pm 0.3\degree$). This means that requiring such a planet to be non-transiting does not necessarily imply that it must also have a large mutual inclination with \sysI b. For $P = 35$ days and $e = 0$, $K < 5$ \mps corresponds to $M_\mathrm{p} \sin i_\mathrm{p} \lesssim 26$ \mearth (at $P = 70$ days and all else equal, the limit is $M_\mathrm{p} \sin i_\mathrm{p} \lesssim 32$ \mearth). Assuming near-coplanarity between any outer non-transiting companions and planet b, these upper limits are consistent with the masses of sub-Neptune- and Neptune-sized planets.

\subsection{Stellar activity} \label{sub:activity}
Stellar activity is known to confound planet detection with RVs, especially when planetary orbits coincide with the stellar rotation period or its harmonics \citep[e.g.,][]{vanderburg16}, Untreated stellar activity signals in RV time series can also lead to mass measurement inaccuracy \citep{rajpaul17}. As such, we investigated \sysI's activity signatures before adopting a final RV model.

\sysI is a chromospherically-inactive, slowly-rotating, G3V dwarf \citep{houk99}. Following the methods of \cite{isaacson24}, we measured \logrhk $=$ \logrhkI using our \keckhires iodine-free template spectrum. From the \keckhires template we also found \vsini $< 2$ \kmps using the \texttt{SpecMatch-Synth} tool \citep{specmatchsynth}. This measurement is consistent at the $3\sigma$-level with the value from \luqueXIX of \vsini $= 4.6 \pm 1.0$ \kmps. Assuming the stellar obliquity is zero (i.e., $i_* = i_\mathrm{p} = 88.4 \pm 0.3\degree$) our upper limit on \vsini corresponds to $P_\mathrm{rot} > 27.6$ days, while the value from \luqueXIX translates to $P_\mathrm{rot} = 12.0 \pm 2.6$ days.

To check for photometric variability indicative of star spots, we extracted the \ktwo Campaign 17 light curve using the \texttt{everest} pipeline \citep{everest:luger18}. After removing the in-transit flux, we followed the methods described in \cite{murphy21} to clean the data with a Savitsky-Golay filter \citep{savitzky64}. We did not find any evidence of an obvious stellar rotation period in the cleaned, out-of-transit flux time series, or in a GLS periodogram of the \ktwo data. We note that NASA's Transiting Exoplanet Survey Satellite \citep[\tess;][]{ricker14} has not observed \sysI as of Cycle 7.

\begin{figure}
    \centering
    \includegraphics[width=\columnwidth]{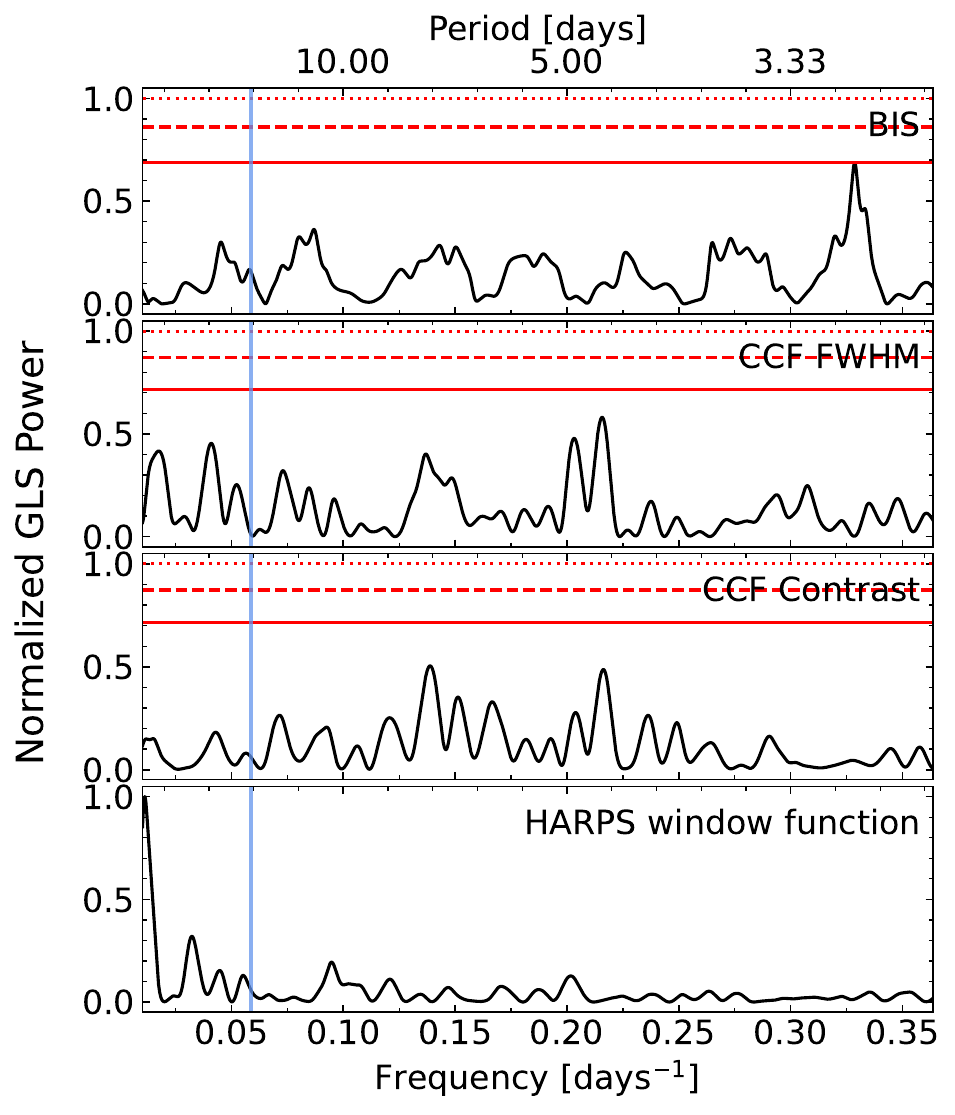}
    \caption{GLS periodograms of the HARPS activity indices and window function. The HARPS data comprise 31 spectra obtained over a baseline of 98 days. The solid blue vertical line corresponds to the orbital period of planet b. The dotted, dashed, and solid red horizontal lines correspond to the 0.1, 1, and 10\% FAP thresholds, respectively, for each periodogram, as computed following \cite{baluev08}.}
    \label{fig:activity_periodograms}
\end{figure}

To check for signs of stellar activity in the spectroscopic data, we examined the HARPS BIS, CCF FWHM, and CCF contrast measurements. The HARPS data was chosen as it represents our largest collection of observations from a single instrument. GLS periodograms of the activity indicator time series are shown in Figure \ref{fig:activity_periodograms}. There are no peaks of significant power in the immediate neighborhood of the transiting planet's orbital period for any of activity indicators.

Despite the lack of an obvious stellar rotation signal in the photometric or spectroscopic data, we added a Gaussian process \citep[GP; e.g.,][]{rasmussen06} to our two-Keplerian RV model in an effort to remove any correlated noise related to activity. We implemented the model with \radvel. For the GP we used the quasi-periodic (QP) kernel \citep[e.g.,][]{grunblatt15, kosiarek21}, which, for each instrument, $i$, quantifies covariance between data observed at times $t$ and $t'$ as
\begin{equation} \label{eqn:qp_kernel}
    k_i (t,t') = \eta_{1,\:i}^2 \ \mathrm{exp} \left[-\frac{(t-t')^2}{\eta_2^2}-\frac{\sin^2(\frac{\pi(t-t')}{\eta_3})}{2 \eta_4^2}\right].
\end{equation}
$\eta_{\text{1-4}}$ are the hyperparameters: $\eta_{1,\:i}$ represents the amplitude of the covariance for instrument $i$, $\eta_2$ is interpreted as the evolutionary timescale of active stellar regions, $\eta_3$ is interpreted as the stellar rotation period, and $\eta_4$ is the length scale of the covariance's periodicity. The hyperparameters are shared between instruments save for the amplitudes, $\eta_{1,\:i}$. Before adding the GP to a model of the RVs, we first trained the GP by fitting it to the HARPS BIS values, which seemed to produce the most well-formed posteriors on the hyperparameters compared to the other activity indicators. The posteriors of $\eta_2$, $\eta_3$, and $\eta_4$ resulting from the training were then used as numerical priors when conducting the MAP fit of the RVs. 

The two-Keplerian plus GP model finds $K_\mathrm{b} = 2.3$ \mps, which matches the result of the Keplerian-only fit ($K_\mathrm{b} = 2.2$ \mps). For the periodic time scale of the GP, the MAP fitting finds $\eta_3 = 23.8$ days, which is about twice the $P_\mathrm{rot}$ estimate implied by the \vsini value from \luqueXIX. While this may seem to suggest that the GP has found the stellar rotation signal (or one of its aliases), the semi-amplitude of planet b remains agnostic. The GP model, with its seven additional parameters (21 free parameters in total), is ruled out by the AICc ($\Delta$AICc $= 18$) when compared to the Keplerian-only model. GPs are also prone to overfitting \citep{blunt23}, so while a stellar activity signal may be present in the existing data, more observations are required before a GP-enabled analysis is warranted. In any case, the inclusion of a GP does not affect the semi-amplitude of planet b.

\subsection{Adopted RV model and posterior estimates}
\begin{figure*}
    \centering
    \includegraphics[width=0.8\textwidth]{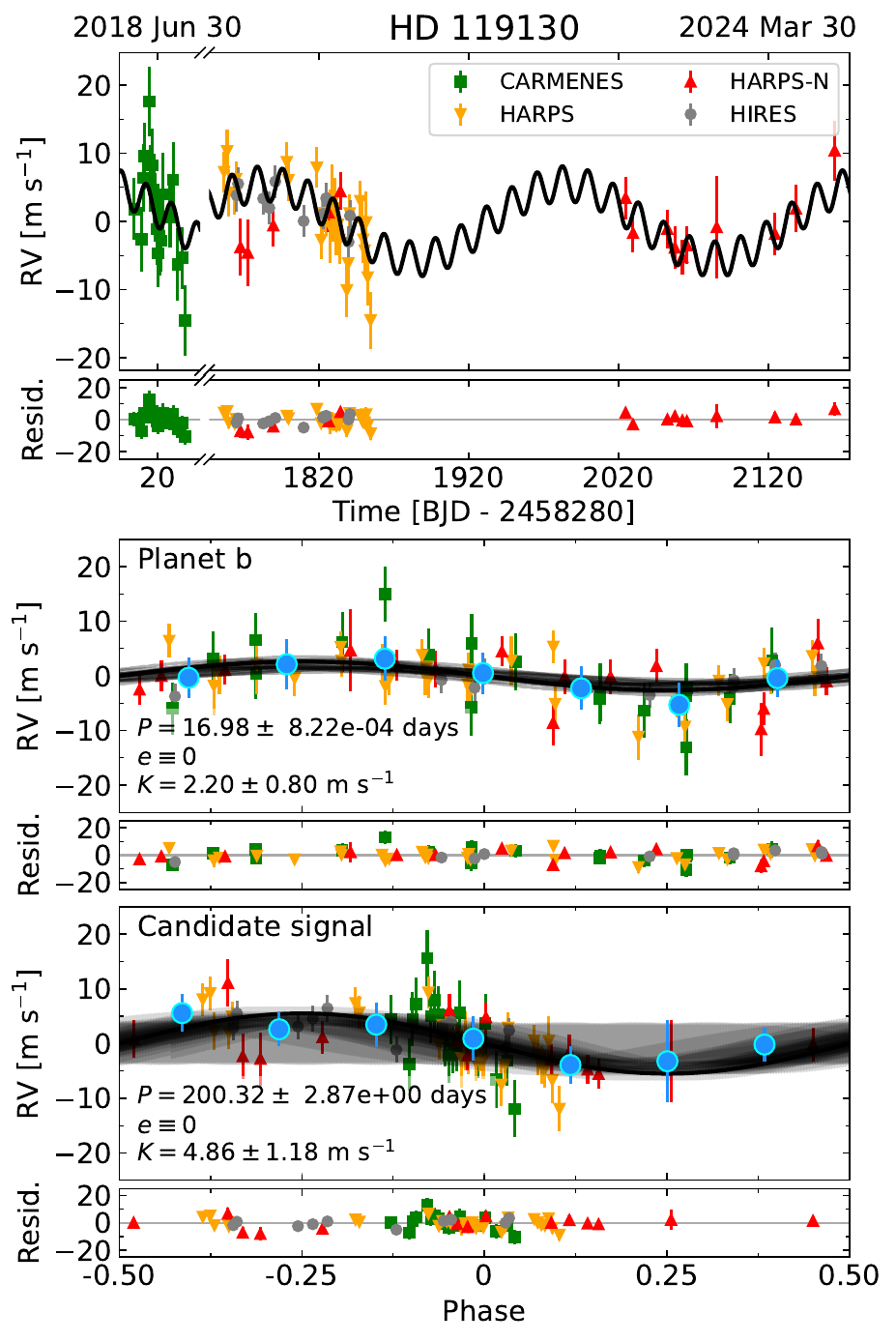}
    \caption{The \sysI RVs. \emph{Top:} The RV time series (data points with error bars), MAP model (solid black line), and residuals about the MAP solution (lower panel). Note the broken x-axis. \emph{Middle:} The RVs folded to the orbital period of \sysI b. RVs binned in units of 0.125 orbital phase are shown as the blue data points. Black lines represent 25 random realizations of the model posterior. Error bars shown represent point wise measurement errors added in quadrature with the associated instrument-specific jitter value that was fit to the data. \emph{Bottom:} The same as above, but for signal c.}
    \label{fig:hd119130_circ_model}
\end{figure*}

To produce our final measurements of the orbit and mass of \sysI b, we fit the full RV data set in \exoplanet using our baseline, circular, two-Keplerian model (Table \ref{tab:hd119130_model}). Observations from the same telescope falling within 8 hours of each other were binned together, resulting in 29 binned HARPS RVs (compared to 31 unbinned RVs) and 15 binned HARPS-N RVs (compared to 16) for the model. The adopted RV model is shown in Figure \ref{fig:hd119130_circ_model} and system parameters are summarized in Table \ref{tab:hd119130_properties}.

To estimate the posterior distributions of the model parameters we used a No-U-Turn Sampler \citep[NUTS;][]{nuts:hoffman14}, an adaptive form of Hamiltonian Monte Carlo \citep[HMC;][]{hmc:duane87, neal12}, implemented with \texttt{pymc3} \citep{pymc3}. The NUTS ran 8 parallel chains with each chain taking 8,000 tuning steps before drawing 2,000 samples. Samples drawn during the tuning period were discarded, similar to how various Markov Chain Monte Carlo (MCMC) methods discard burn-in samples. The chains were concatenated to produce a total of 16,000 samples from the posterior of each model parameter. We computed the rank-normalized Gelman-Rubin statistic \citep[$\hat{R}$;][]{gelman92, vehtari21} for the chains and find $\hat{R} < 1.001$ for all model parameters, suggesting convergence. We also computed the rank-normalized effective sample sizes of the bulk and tails of each posterior distribution \citep{vehtari21} and find that they are all sufficiently large ($N_\mathrm{eff} > 1000$) compared to the recommended threshold ($N_\mathrm{eff} = 400$).

\begin{deluxetable}{lcr}
\tablecaption{\sysI system properties \label{tab:hd119130_properties}}
\tabletypesize{\small}
\startdata
\tablehead{
    \colhead{Parameter [units]} & \colhead{Value} & \colhead{Provenance}
}\\
\multicolumn{3}{l}{\emph{System information}} \\
EPIC ID & 212628254 & EPIC \\
TIC ID  & 124573851 & Guerrero \\
R.A. [J2000] & 13:41:30.2 & \gaiadrthree \\
decl. [J2000] & $-$09:56:45.9 & \gaiadrthree \\
Parallax [mas] & $8.81 \pm 0.02$ & \gaiadrthree \\
$V$ [mag] & $9.92 \pm 0.01$ & Munari \\
$J$ [mag] & $8.73 \pm 0.02$ & \twomass \\
\vspace{1pt} \\
\multicolumn{3}{l}{\emph{Stellar parameters}} \\
\teff [K] & $5725 \pm 65$ & \luqueXIX \\
$\log_{10} g$ & $4.30 \pm 0.15$ & \luqueXIX \\
\vsini [km s$^{-1}$] & $< 2$ & This work\\ 
\vsini [km s$^{-1}$] & $4.6 \pm 1.0$ & \luqueXIX \\ 
\mstar [\msun] & $1.00 \pm 0.03$ & \luqueXIX \\
\rstar [\rsun] & $1.09 \pm 0.03$ & \luqueXIX \\ 
\logrhk & \logrhkI & This work \\
\vspace{1pt} \\
\multicolumn{3}{l}{\emph{Planet b measured quantities}} \\
$P$ [days] & $16.9841 \pm 0.0008$ & \luqueXIX \\
$T_\mathrm{c}$ [BJD] & \tcIb & This work \\
$K$ [m s$^{-1}$] & \KIb & This work \\
$e$ & $\equiv 0$ & This work \\
$\omega_\mathrm{p}$ & \nodata & This work \\
\vspace{1pt} \\
\multicolumn{3}{l}{\emph{Planet b derived quantities}} \\
$a$ [au] & $0.13 \pm 0.01$ & \luqueXIX \\
\teq [K] & $795^{+33}_{-28}$ & \luqueXIX \\
$T_\mathrm{dur}$ [hr] & $3.66^{+0.07}_{-0.08}$ & \luqueXIX \\
$i_\mathrm{p}$ [deg] & $88.4^{+0.2}_{-0.3}$ & \luqueXIX \\
\rplanet [$R_\mathrm{\oplus}$] & \rpIb & \luqueXIX \\
\mplanet [$M_\mathrm{\oplus}$] & \mpIb & This work \\
$\rho_\mathrm{p}$ [g cm$^{-3}$] & \rhoIb & This work \\
\vspace{1pt} \\
\multicolumn{3}{l}{\emph{Instrument parameters}} \\
$\gamma_\mathrm{CARMENES}$ [m s$^{-1}$] & \gammarvCARMENESI & This work \\
$\sigma_\mathrm{CARMENES}$ [m s$^{-1}$] & \sigmarvCARMENESI & This work  \\
$\gamma_\mathrm{HARPS}$ [m s$^{-1}$]    & \gammarvHARPSI  & This work \\
$\sigma_\mathrm{HARPS}$ [m s$^{-1}$]    & \sigmarvHARPSI & This work \\
$\gamma_\mathrm{HARPS-N}$ [m s$^{-1}$]  & \gammarvHARPSNI  & This work \\
$\sigma_\mathrm{HARPS-N}$ [m s$^{-1}$]  & \sigmarvHARPSNI & This work \\
$\gamma_\mathrm{HIRES}$ [m s$^{-1}$]    & \gammarvHIRESI  & This work \\
$\sigma_\mathrm{HIRES}$ [m s$^{-1}$]    & \sigmarvHIRESI & This work
\enddata
\tablecomments{\teq calculated assuming zero Bond albedo and perfect day-night heat redistribution. References for provenance values: EPIC \citep{epic}, Guerrero \citep{guerrero21}, \gaiadrthree \citep{gaia, gaiadr3}, Munari \citep{munari14}, and \twomass \citep{2mass}.}
\end{deluxetable}

\section{Understanding the Mass Measurement Inaccuracy} \label{sec:explanations}

\subsection{Sanity checks}\label{sec:sanity_checks}
\begin{figure}
    \centering
    \includegraphics[width=\columnwidth]{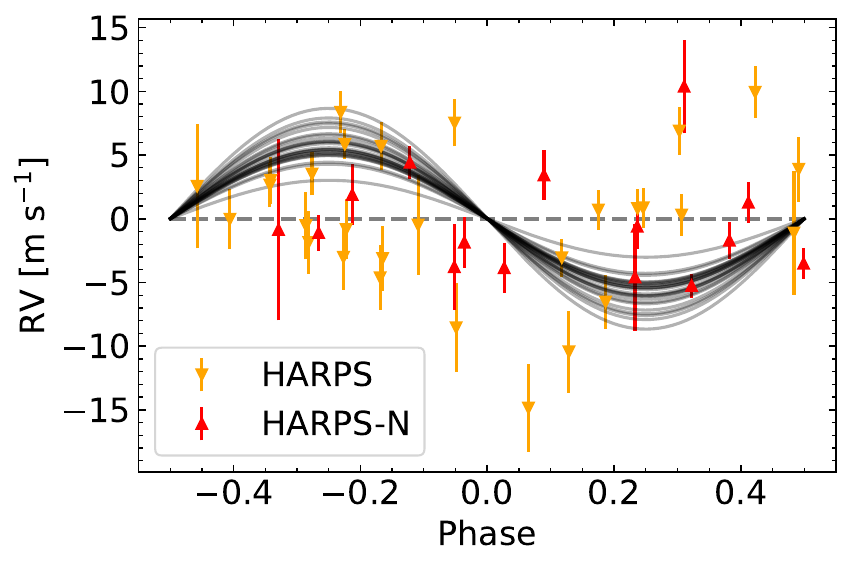}
    \caption{The HARPS and HARPS-N data (with instrumental offsets applied) compared to 25 random realizations of \sysI b's orbit, generated using the $P$, \transitTime, and $K$ values from \luqueXIX. The new data are inconsistent with the \luqueXIX solution.}
    \label{fig:l19_orbit_comparison}
\end{figure}

Why is the \luqueXIX mass measurement inaccurate? Before exploring possible explanations, we conduct two brief sanity checks to further convince the reader that the discrepancy between our measurement and the \luqueXIX value cannot be explained by systematic differences in our analysis. First, we compared the HARPS and HARPS-N data---which have a similar sampling cadence compared to the CARMENES RVs---to the orbital solution from \luqueXIX. It is clear by eye that the new data do not agree with the orbit (see Figure \ref{fig:l19_orbit_comparison}). The RMS of the HARPS and HARPS-N data about the \luqueXIX orbit is 7.3 \mps, compared to 4.0 \mps for all of the data about our adopted solution.

\begin{figure}
    \centering
    \includegraphics[width=\columnwidth]{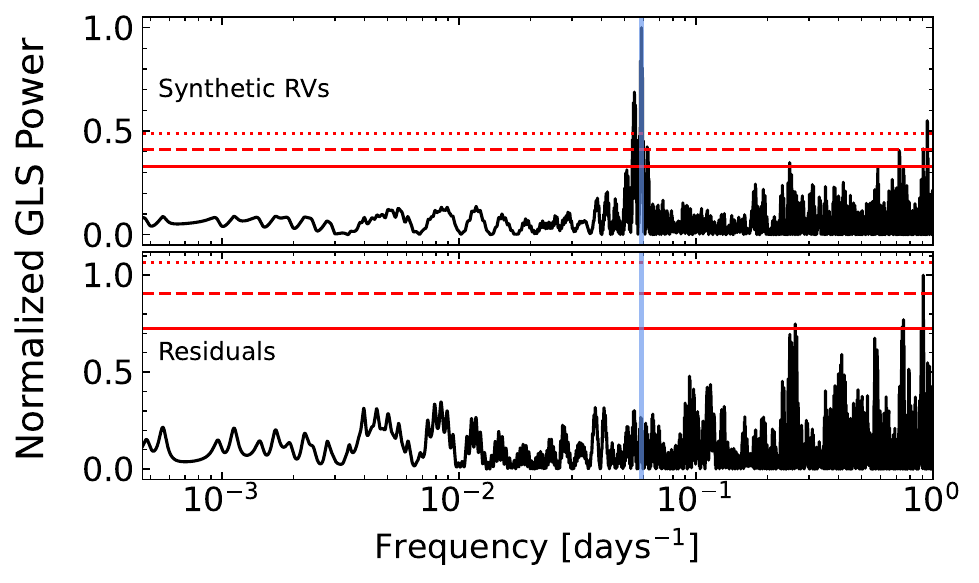}
    \caption{\emph{Top:} GLS periodogram of a synthetic RV time series generated by assuming the \luqueXIX orbital solution for \sysI b and producing an RV at each observation time stamp in our real data set. We assume typical measurement errors and white noise. The vertical blue line marks the orbital period of \sysI b. As before, the horizontal lines correspond to the 10\%, 1\%, and 0.1\% FAP thresholds. If the planet did indeed have $K_\mathrm{b} = 6$ \mps, it would have been easily detected in our data, assuming it is the only periodic signal present. \emph{Bottom:} The residuals of the synthetic RVs after removing the \luqueXIX orbit for planet b.}
    \label{fig:synth_periodograms}
\end{figure}

Second, if \sysI b did indeed have $K_\mathrm{b} = 6$ \mps, the signal would have been easily detected in our data given our number of measurements, sampling cadence, observing baseline, and typical measurement uncertainty. Using the \luqueXIX orbital solution, we generated synthetic RV measurements at our observation time stamps with a measurement uncertainty of 2.5 \mps (the average measurement uncertainty of our data) and 3 \mps of additional white noise added in quadrature (the average of the best-fitting RV jitter values across the four instruments). The GLS periodogram of the synthetic time series is shown in the top panel of Figure \ref{fig:synth_periodograms}. In this case, the planet is easily detected with an FAP of much less than 0.1\%. Of course, this simulation assumes that the transiting planet is the only periodic signal present in the data.

\subsection{Observing cadence and number of measurements} \label{sec:resampling}
Is the \luqueXIX mass measurement's inaccuracy due to an unfortunate conspiracy between imperfect sampling cadence, observing baseline, and the number of CARMENES observations? Following the methods of \cite{accuracyMurphy24arxiv}, we resampled the data using different combinations of minimum observing cadence (MOC) and the number of observations in order to visualize how the MAP value of $K_\mathrm{b}$ would change due to these properties alone. 

\begin{figure}
    \centering
    \includegraphics[width=\columnwidth]{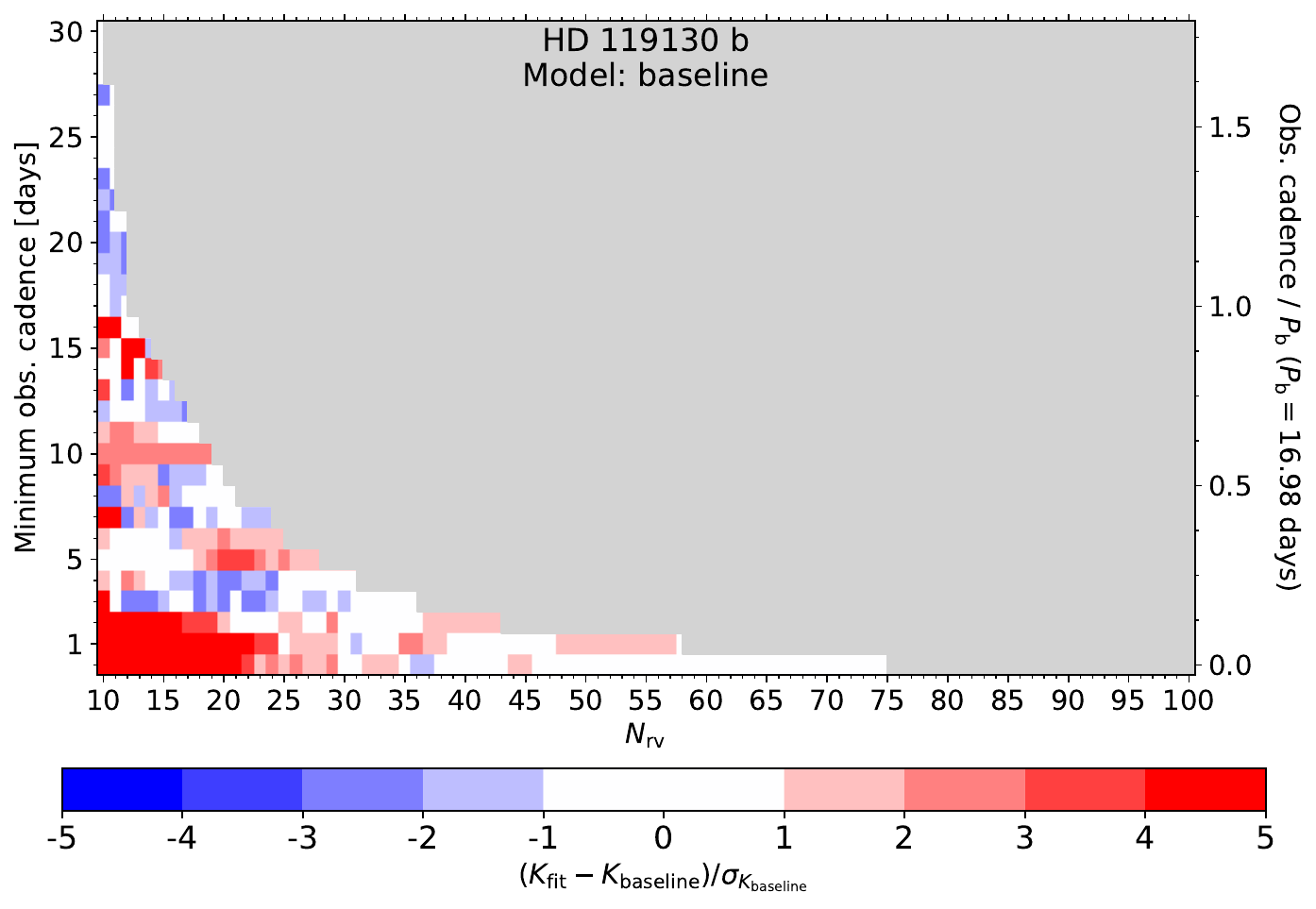}
    \caption{The results of the resampling experiment for \sysI b following the methods of \cite{accuracyMurphy24arxiv}. Each cell represents a resampled realization of the full data set according to specified MOC (left y-axis; i.e., a lower limit enforced on the time between consecutive measurements) and \nrv (x-axis) values. The right y-axis is the same as the left, but it is now shown in units of the orbital period of the planet. The color of each cell represents the difference between the planet's measured semi-amplitude using that cell's resampled data set, $K_\mathrm{fit}$, and the measured semi-amplitude of the planet from the baseline model ($K_\mathrm{baseline}$), in units of the uncertainty on $K_\mathrm{baseline}$. Gray regions represent combinations of minimum observing cadence and \nrv that are not supported by the data. The right-most, non-gray cell along the bottom row represents the original time series, and is therefore in agreement with $K_\mathrm{baseline}$ by construction.}
    \label{fig:hd119130_real_base}
\end{figure}

The results of the resampling experiment are shown in Figure \ref{fig:hd119130_real_base}. For resampled datasets similar in size to the \luqueXIX time series ($N_\mathrm{RV} \approx 20$), $K_\mathrm{b}$ can vary widely depending on the chosen MOC. Perhaps this means that the \luqueXIX overestimate is a stochastic outcome of the interplay between the sampling cadence of their observations, the orbit of \sysI b, and the phase of any additional signals (e.g., stellar activity) present when the CARMENES data were collected.

\subsection{Rotational modulation}
As discussed in \S\ref{sub:activity}, $P \approx 12$ days could plausibly represent the stellar rotation period or its first harmonic (depending on which value of \vsini is adopted). Is it possible that a coherent stellar rotation signal, interfering constructively with the planetary signal over the course of the \luqueXIX observations, was responsible for inflating $K_\mathrm{b}$ from 2.2 \mps to 6.1 \mps? To test this explanation, we conducted simulations following the methods of \cite{rajpaul17}. 

Using Equation 1 in \cite{rajpaul17}, we generated synthetic RVs using the observing cadence, number of observations, and noise level of the CARMENES data. We assumed two sinusoidal signals were present. The first signal corresponded to the RV orbit of planet b:
\begin{equation}
    y_{1,\:i} = K_1 \sin \big(\frac{2 \pi t_i}{P_1}\big) + \mathcal{N}(0, \sigma_{\mathrm{CARMENES}}),
\end{equation}
where $K_1 = 2.2$ \mps, $P_1 = 16.98$ days, $\sigma_{\mathrm{CARMENES}} = 3$ \mps, and $t_i$ is the time stamp of the $i$-th observation (with the first time stamp arbitrarily set to zero). We then generated 10,000 realizations of a second sinusoid, meant to represent a coherent rotation signal:
\begin{equation}
     y_{2,\:i} = K_2 \sin \big(\frac{2 \pi t_i}{P_2} + \phi_2 \big) + \mathcal{N}(0, \sigma_{\mathrm{CARMENES}}),
\end{equation}
where $K_2 = [1.0, 1.04, ..., 4.96]$ \mps (an array of 100 elements evenly-spaced between 1--5 \mps)  and $\phi_2 = 2 \pi \times [0, 0.01, ..., 0.99]$ (an array of 100 elements evenly-spaced between 0--$2\pi$ radians). 

\begin{figure}
    \centering
    \includegraphics[width=\columnwidth]{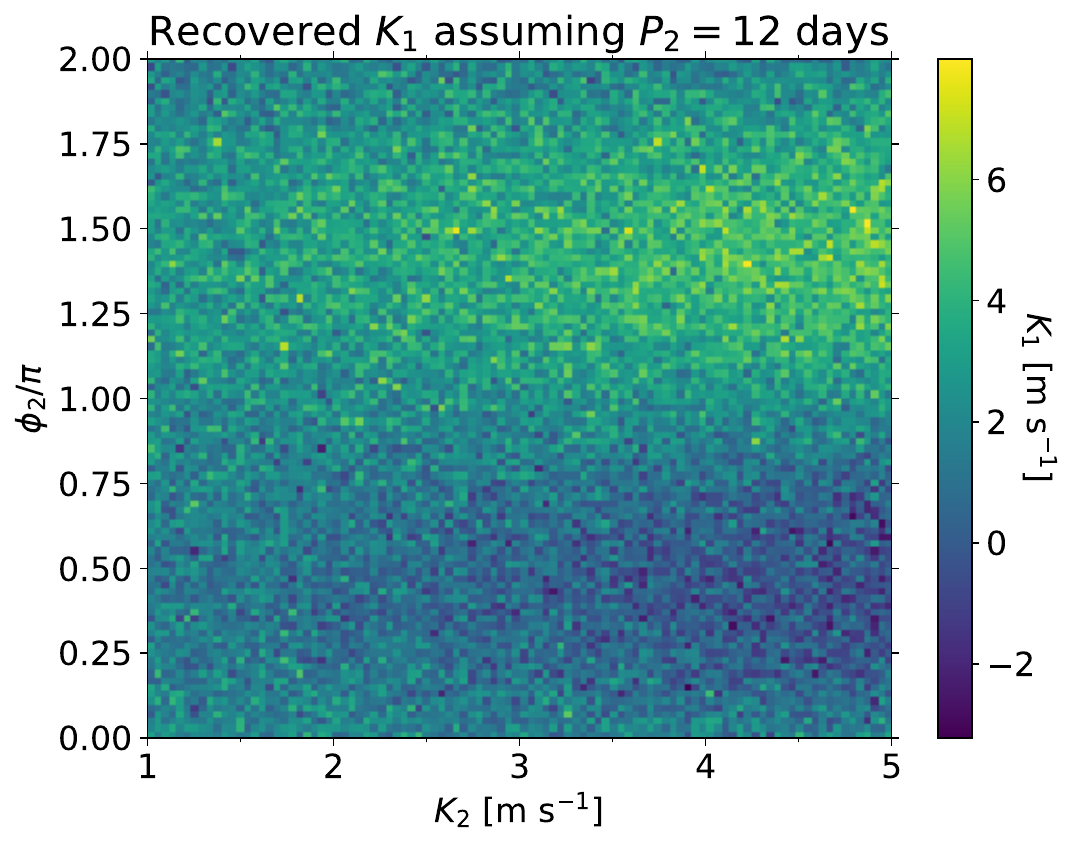}
    \caption{A map of best-fitting $K_1$ values fit using a one-planet model, but where the (synthetic) data contain a second signal at $P_2 = 12$ days. The second signal is specified by the different combinations of $K_2$ and $\phi_2$. We see that in certain cases, $K_1$ is inflated to values similar to the result from \luqueXIX.}
    \label{fig:rajpaul_experiment}
\end{figure}

For each of the 10,000 realizations of the synthetic data, $y = y_1 + y_2$, we fit the sinusoid $y = K \sin \big(\frac{2 \pi t}{P}\big)$, where $P = 16.98$ days. We recorded the best-fitting $K$ value for each combination of $K_2$ and $\phi_2$. The results are shown in Figure \ref{fig:rajpaul_experiment}. We see that there are indeed instances where the best-fitting $K_1$ value is $\geq 6$ \mps, though they only account for $\sim1\%$ of all trials. $K_1 \geq 3.9$ \mps (i.e., $K_1 \geq$ the \luqueXIX value minus $2\sigma$) accounts for $15\%$ of all trials. 

From these statistics alone, the rotational modulation explanation may seem a bit unlikely. But, of course, this experiment was only conducted for one choice of $P_2$. In any case, our simple experiment demonstrates that it is at least feasible for a coherent stellar rotation signal on the order of a few \mps to have artificially inflated the RV semi-amplitude of \sysI b.

\section{Planetary composition, formation, and evolution} \label{sec:discussion}
\begin{figure}
    \centering
    \includegraphics[width=\columnwidth]{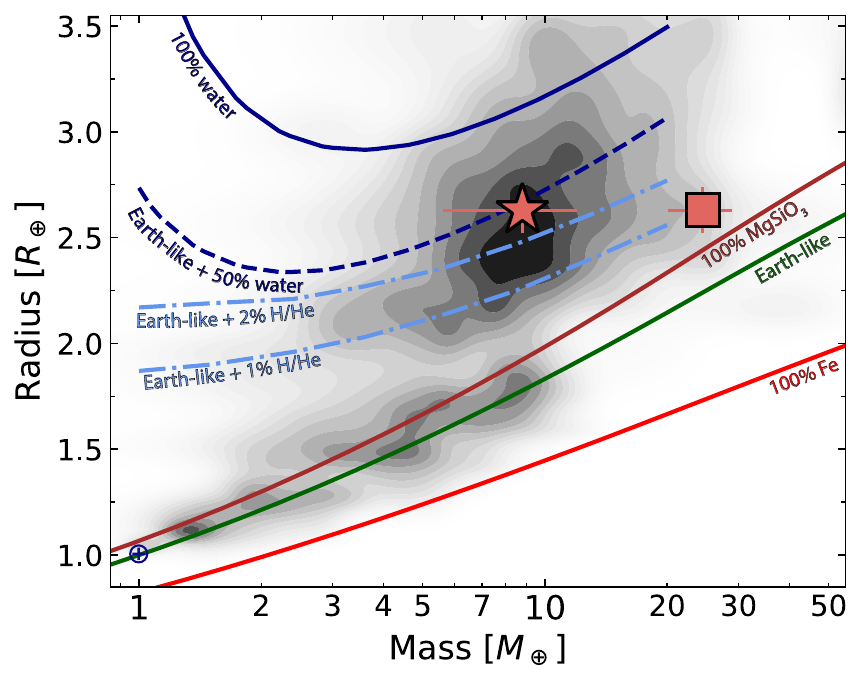}
    \caption{The mass-radius diagram in the sub-Neptune and super-Earth regime. Contours are computed using Gaussian kernel density estimation on the population of confirmed planets with better than 33\% and 15\% measurement precision on mass and radius, respectively \citep{nea}. Composition curves are shown for reference. The two dash-dot light blue curves represent Earth-like cores surrounded by 1\% and 2\% by mass H/He-dominated, solar-metallicity envelopes for a planet receiving 10 times Earth's insolation orbiting a 10 Gyr-old star \citep{lopezforney14}. \sysI b's instellation is $S_\mathrm{p} = 67$ $S_\mathrm{\oplus}$. The solid red, green, and brown curves represent a pure iron, Earth-like, and pure silicate composition, respectively \citep{zeng16}. The dashed dark blue curve represents an irradiated ``water world'' at \teq $= 800$ K made of an Earth-like core and 50\% \water, by mass, in the form of a supercritical \water fluid layer and a steam-dominated envelope; the solid dark blue curve represents a composition of 100\% supercritical \water and steam at \teq $= 800$ K \citep{aguichine21}. \sysI b is shown as both the square (\luqueXIX mass measurement) and the star (this work) for comparison.}
    \label{fig:mass_radius}
\end{figure}

The mass-radius diagram shown in Figure \ref{fig:mass_radius} includes a visual comparison of the drastic change in the mass estimate of \sysI b between \luqueXIX and this work. How does this new mass measurement change the interpretation of the planet's bulk composition and formation and evolution history?

Given our relatively imprecise ($2.75\sigma$) constraint on the mass of \sysI b, we forgo a detailed analysis of the planet's interior structure. We note, however, that the new measurement is inconsistent with a pure silicate composition, as previously suggested by \luqueXIX. The planet's revised bulk density (\rhoIb \gcc) is consistent with a planet comprising an Earth-like core beneath a substantial (1--2\% by mass), H/He-dominated volatile envelope or a water-rich core beneath a high-metallicity atmosphere.

Using Equation 15 from \cite{lecavelier07}, \sysI b could lose up to 0.04 \mearth over 10 Gyr due to extreme ultraviolet radiation from its G dwarf host at its current orbit. According to the grid from \cite{lopezforney14}, this would translate to a reduction in the planet's radius by $\sim$0.2 \rearth. However, even at its lower mass, \sysI b would still require a disk mass enhancement factor of $\sim$30 above the minimum-mass solar nebula to form at $a = 0.13$ au \citep[see Equation 7 from][]{schlichting14}, making in situ formation implausible and meaning that the mass loss predicted above is likely an overestimate. As \luqueXIX concluded, \sysI b must have formed farther out in the disk and subsequently migrated inward due to dynamical interactions with either the primordial gaseous protoplanetary disk or other bodies.

The possibility that \sysI b formed farther out in the disk---perhaps beyond the snow line, where large planetary embryos are expected to grow quickly \citep{kokubo02, morbidelli15}---does not necessarily preclude a water-poor composition \citep{raymond18}. Measurements of the planet's atmospheric metallicity could break the degeneracy in composition by determining if the planet has a H/He- or steam-dominated envelope. The planet's Transmission Spectroscopy Metric \citep[TSM;][]{kempton18}, a \jwst \snr proxy, is $\mathrm{TSM} = 31^{+19}_{-9}$, which corresponds to the 54th percentile of TSM values for the confirmed planets that underlie the contours in Figure \ref{fig:mass_radius}.

\section{Conclusion} \label{sec:conclusion}
We obtained 57 new RVs of the G dwarf, \sysI, using HARPS, HARPS-N, and HIRES. Combined with the 18 existing CARMENES RVs from \luqueXIX, our follow-up observations reveal that the transiting sub-Neptune, \sysI b, is not unusually massive for its size, as previously thought. In fact, our mass measurement of \sysI b is almost three times smaller than the value from \luqueXIX. The case of \sysI b serves as a cautionary tale for observers analyzing modestly-sized RV data sets. 

Our conclusions are summarized as follows:
\begin{itemize}
    \item \sysI b (\rplanet $=$ \rpIb \rearth) has \mplanet $=$ \mpIb \mearth (\mplanet $<$ \mpIbupperlim \mearth at 98\% confidence). Our mass measurement differs from the \luqueXIX value at the $2.9\sigma$-level. The planet is not an ``ultra-dense'' sub-Neptune.
    
    \item We detect a candidate signal in the RVs at $P \approx 200$ days with $K =$ \KIc \mps. Additional RV monitoring is required to determine the signal's true nature.

    \item The inaccuracy of the \luqueXIX mass measurement can be explained by the presence of an additional sinusoidal signal in the data. With $P = 12$ days, this signal could feasibly represent either the stellar rotation period or its first harmonic.
\end{itemize}

As we begin to study small-planet demographics in the mass-radius plane \citep[e.g.,][]{luque22}, the community should endeavor to ensure that these mass measurements are both precise \emph{and} accurate, lest they bias our understanding of the physics of planet formation.

\vspace{0.5cm}
\begin{center}
ACKNOWLEDGMENTS
\end{center}

The authors thank the anonymous reviewer for their thoughtful report, which improved the manuscript.

The authors wish to recognize and acknowledge the very significant cultural role and reverence that Maunakea has always had within the indigenous Hawaiian community. We are most fortunate to have the opportunity to conduct observations from this sacred mountain, which is now colonized land.

J.M.A.M. thanks Jacob Bean for detailed comments that improved the manuscript. J.M.A.M. thanks the \keckhires observers for obtaining data of \sysI. In order of the number of observations contributed, the observers are: Judah Van Zandt, Fei Dai, Steven Giacalone, Rae Holcomb, Howard Isaacson, Corey Beard, Samuel Yee, Lauren Weiss, Jingwen Zhang, and Luke Bouma.

J.M.A.M. is supported by the National Science Foundation (NSF) Graduate Research Fellowship Program (GRFP) under Grant No. DGE-1842400. J.M.A.M. and N.M.B. acknowledge support from NASA’S Interdisciplinary Consortia for Astrobiology Research (NNH19ZDA001N-ICAR) under award number 19-ICAR19\_2-0041.

V.V.E. is supported by UK's Science \& Technology Facilities Council (STFC) through the STFC grants ST/W001136/1 and ST/S000216/1.

Support for this work was provided by NASA through the NASA Hubble Fellowship grant \#HF2-51559 awarded by the Space Telescope Science Institute, which is operated by the Association of Universities for Research in Astronomy, Inc., for NASA, under contract NAS5-26555.

This work used \texttt{Expanse} at the San Diego Supercomputer Center through allocation PHY220015 from the Advanced Cyberinfrastructure Coordination Ecosystem: Services \& Support (ACCESS) program, which is supported by NSF grants 2138259, 2138286, 2138307, 2137603, and 2138296.

Some of the data presented herein were obtained at Keck Observatory, which is a private 501(c)3 non-profit organization operated as a scientific partnership among the California Institute of Technology, the University of California, and the National Aeronautics and Space Administration. The Observatory was made possible by the generous financial support of the W. M. Keck Foundation.

We acknowledge financial support from the Agencia Estatal de Investigaci\'on of the Ministerio de Ciencia e Innovaci\'on MCIN/AEI/10.13039/501100011033 and the ERDF ``A way of making Europe'' through project PID2021-125627OB-C32, and from the Centre of Excellence ``Severo Ochoa'' award to the Instituto de Astrofisica de Canarias.

G.N. acknowledges financial support from the Ministry of Science and Higher Education program the "Excellence Initiative - Research University" conducted at the Centre of Excellence in Astrophysics and Astrochemistry of the Nicolaus Copernicus University in Toru\'n, Poland. G.N. gratefully acknowledges the Centre of Informatics Tricity Academic Supercomputer and networK (CI TASK, Gda\'nsk, Poland) for computing resources (grant no. PT01016).

This work benefited from the 2023 Exoplanet Summer Program in the Other Worlds Laboratory (OWL) at the University of California, Santa Cruz, a program funded by the Heising-Simons Foundation.

\facilities{HARPS, HARPS-N, Keck:I (HIRES).}

\software{\texttt{exoplanet} \citep{exoplanet:exoplanet}, \texttt{matplotlib} \citep{matplotlib}, \texttt{numpy} \citep{numpy}, \texttt{pandas} \citep{pandas}, Python 3 \citep{python3}, \radvel \citep{radvel}, \rvsearch \citep{rosenthal21}.}

\bibliography{main}{}

\begin{thebibliography}{}
\expandafter\ifx\csname natexlab\endcsname\relax\def\natexlab#1{#1}\fi
\providecommand{\url}[1]{\href{#1}{#1}}
\providecommand{\dodoi}[1]{doi:~\href{http://doi.org/#1}{\nolinkurl{#1}}}
\providecommand{\doeprint}[1]{\href{http://ascl.net/#1}{\nolinkurl{http://ascl.net/#1}}}
\providecommand{\doarXiv}[1]{\href{https://arxiv.org/abs/#1}{\nolinkurl{https://arxiv.org/abs/#1}}}

\bibitem[{{Aguichine} {et~al.}(2021){Aguichine}, {Mousis}, {Deleuil}, \& {Marcq}}]{aguichine21}
{Aguichine}, A., {Mousis}, O., {Deleuil}, M., \& {Marcq}, E. 2021, \apj, 914, 84, \dodoi{10.3847/1538-4357/abfa99}

\bibitem[{{Akaike}(1974)}]{akaike74}
{Akaike}, H. 1974, IEEE Transactions on Automatic Control, 19, 716

\bibitem[{{Akana Murphy} {et~al.}(2024){Akana Murphy}, {Luque}, \& {Batalha}}]{accuracyMurphy24arxiv}
{Akana Murphy}, J.~M., {Luque}, R., \& {Batalha}, N.~M. 2024, arXiv e-prints, arXiv:2411.02521, \dodoi{10.48550/arXiv.2411.02521}

\bibitem[{{Akana Murphy} {et~al.}(2021){Akana Murphy}, {Kosiarek}, {Batalha}, {Gonzales}, {Isaacson}, {Petigura}, {Weiss}, {Grunblatt}, {Ciardi}, {Fulton}, {Hirsch}, {Behmard}, \& {Rosenthal}}]{murphy21}
{Akana Murphy}, J.~M., {Kosiarek}, M.~R., {Batalha}, N.~M., {et~al.} 2021, \aj, 162, 294, \dodoi{10.3847/1538-3881/ac2830}

\bibitem[{{Armstrong} {et~al.}(2020){Armstrong}, {Lopez}, {Adibekyan}, {Booth}, {Bryant}, {Collins}, {Deleuil}, {Emsenhuber}, {Huang}, {King}, {Lillo-Box}, {Lissauer}, {Matthews}, {Mousis}, {Nielsen}, {Osborn}, {Otegi}, {Santos}, {Sousa}, {Stassun}, {Veras}, {Ziegler}, {Acton}, {Almenara}, {Anderson}, {Barrado}, {Barros}, {Bayliss}, {Belardi}, {Bouchy}, {Brice{\~n}o}, {Brogi}, {Brown}, {Burleigh}, {Casewell}, {Chaushev}, {Ciardi}, {Collins}, {Col{\'o}n}, {Cooke}, {Crossfield}, {D{\'\i}az}, {Delgado Mena}, {Demangeon}, {Dorn}, {Dumusque}, {Eigm{\"u}ller}, {Fausnaugh}, {Figueira}, {Gan}, {Gandhi}, {Gill}, {Gonzales}, {Goad}, {G{\"u}nther}, {Helled}, {Hojjatpanah}, {Howell}, {Jackman}, {Jenkins}, {Jenkins}, {Jensen}, {Kennedy}, {Latham}, {Law}, {Lendl}, {Lozovsky}, {Mann}, {Moyano}, {McCormac}, {Meru}, {Mordasini}, {Osborn}, {Pollacco}, {Queloz}, {Raynard}, {Ricker}, {Rowden}, {Santerne}, {Schlieder}, {Seager}, {Sha}, {Tan}, {Tilbrook}, {Ting}, {Udry}, {Vanderspek}, {Watson}, {West}, {Wilson}, {Winn},
  {Wheatley}, {Villasenor}, {Vines}, \& {Zhan}}]{armstrong20}
{Armstrong}, D.~J., {Lopez}, T.~A., {Adibekyan}, V., {et~al.} 2020, \nat, 583, 39, \dodoi{10.1038/s41586-020-2421-7}

\bibitem[{{Baluev}(2008)}]{baluev08}
{Baluev}, R.~V. 2008, \mnras, 385, 1279, \dodoi{10.1111/j.1365-2966.2008.12689.x}

\bibitem[{Betancourt \& Girolami(2015)}]{betancourt15}
Betancourt, M., \& Girolami, M. 2015, Current trends in Bayesian methodology with applications, 79, 2.
\newblock \doarXiv{1312.0906}

\bibitem[{{Blunt} {et~al.}(2023){Blunt}, {Carvalho}, {David}, {Beichman}, {Zink}, {Gaidos}, {Behmard}, {Bouma}, {Cody}, {Dai}, {Foreman-Mackey}, {Grunblatt}, {Howard}, {Kosiarek}, {Knutson}, {Rubenzahl}, {Beard}, {Chontos}, {Giacalone}, {Hirano}, {Johnson}, {Lubin}, {Akana Murphy}, {Petigura}, {Van Zandt}, \& {Weiss}}]{blunt23}
{Blunt}, S., {Carvalho}, A., {David}, T.~J., {et~al.} 2023, \aj, 166, 62, \dodoi{10.3847/1538-3881/acde78}

\bibitem[{{Burnham} \& {Anderson}(2002)}]{burnham02}
{Burnham}, K.~P., \& {Anderson}, D.~R. 2002, Model Selection and Multimodel Inference (Springer-Verlag New York)

\bibitem[{Burnham \& Anderson(2004)}]{burnham04}
Burnham, K.~P., \& Anderson, D.~R. 2004, Sociological Methods \& Research, 33, 261, \dodoi{10.1177/0049124104268644}

\bibitem[{{Burt} {et~al.}(2018){Burt}, {Holden}, {Wolfgang}, \& {Bouma}}]{burt18}
{Burt}, J., {Holden}, B., {Wolfgang}, A., \& {Bouma}, L.~G. 2018, \aj, 156, 255, \dodoi{10.3847/1538-3881/aae697}

\bibitem[{Butler {et~al.}(1996)Butler, Marcy, Williams, McCarthy, Dosanjh, \& Vogt}]{butler96}
Butler, R.~P., Marcy, G.~W., Williams, E., {et~al.} 1996, Publications of the Astronomical Society of the Pacific, 108, 500, \dodoi{10.1086/133755}

\bibitem[{{Cosentino} {et~al.}(2012){Cosentino}, {Lovis}, {Pepe}, {Collier Cameron}, {Latham}, {Molinari}, {Udry}, {Bezawada}, {Black}, {Born}, {Buchschacher}, {Charbonneau}, {Figueira}, {Fleury}, {Galli}, {Gallie}, {Gao}, {Ghedina}, {Gonzalez}, {Gonzalez}, {Guerra}, {Henry}, {Horne}, {Hughes}, {Kelly}, {Lodi}, {Lunney}, {Maire}, {Mayor}, {Micela}, {Ordway}, {Peacock}, {Phillips}, {Piotto}, {Pollacco}, {Queloz}, {Rice}, {Riverol}, {Riverol}, {San Juan}, {Sasselov}, {Segransan}, {Sozzetti}, {Sosnowska}, {Stobie}, {Szentgyorgyi}, {Vick}, \& {Weber}}]{harpsn}
{Cosentino}, R., {Lovis}, C., {Pepe}, F., {et~al.} 2012, in Society of Photo-Optical Instrumentation Engineers (SPIE) Conference Series, Vol. 8446, Ground-based and Airborne Instrumentation for Astronomy IV, ed. I.~S. {McLean}, S.~K. {Ramsay}, \& H.~{Takami}, 84461V, \dodoi{10.1117/12.925738}

\bibitem[{{Cosentino} {et~al.}(2014){Cosentino}, {Lovis}, {Pepe}, {Collier Cameron}, {Latham}, {Molinari}, {Udry}, {Bezawada}, {Buchschacher}, {Figueira}, {Fleury}, {Ghedina}, {Glenday}, {Gonzalez}, {Guerra}, {Henry}, {Hughes}, {Maire}, {Motalebi}, \& {Phillips}}]{cosentino14}
{Cosentino}, R., {Lovis}, C., {Pepe}, F., {et~al.} 2014, in Society of Photo-Optical Instrumentation Engineers (SPIE) Conference Series, Vol. 9147, Ground-based and Airborne Instrumentation for Astronomy V, ed. S.~K. {Ramsay}, I.~S. {McLean}, \& H.~{Takami}, 91478C, \dodoi{10.1117/12.2055813}

\bibitem[{{Dawson} \& {Chiang}(2014)}]{dawsonChiang14}
{Dawson}, R.~I., \& {Chiang}, E. 2014, Science, 346, 212, \dodoi{10.1126/science.1256943}

\bibitem[{{Dragomir} {et~al.}(2019){Dragomir}, {Teske}, {G{\"u}nther}, {S{\'e}gransan}, {Burt}, {Huang}, {Vanderburg}, {Matthews}, {Dumusque}, {Stassun}, {Pepper}, {Ricker}, {Vanderspek}, {Latham}, {Seager}, {Winn}, {Jenkins}, {Beatty}, {Bouchy}, {Brown}, {Butler}, {Ciardi}, {Crane}, {Eastman}, {Fossati}, {Francis}, {Fulton}, {Gaudi}, {Goeke}, {James}, {Klaus}, {Kuhn}, {Lovis}, {Lund}, {McDermott}, {Paegert}, {Pepe}, {Rodriguez}, {Sha}, {Shectman}, {Shporer}, {Siverd}, {Garcia Soto}, {Stevens}, {Twicken}, {Udry}, {Villanueva}, {Wang}, {Wohler}, {Yao}, \& {Zhan}}]{dragomir19}
{Dragomir}, D., {Teske}, J., {G{\"u}nther}, M.~N., {et~al.} 2019, \apjl, 875, L7, \dodoi{10.3847/2041-8213/ab12ed}

\bibitem[{{Duane} {et~al.}(1987){Duane}, {Kennedy}, {Pendleton}, \& {Roweth}}]{hmc:duane87}
{Duane}, S., {Kennedy}, A.~D., {Pendleton}, B.~J., \& {Roweth}, D. 1987, Physics Letters B, 195, 216, \dodoi{10.1016/0370-2693(87)91197-X}

\bibitem[{{Eastman} {et~al.}(2010){Eastman}, {Siverd}, \& {Gaudi}}]{eastman10}
{Eastman}, J., {Siverd}, R., \& {Gaudi}, B.~S. 2010, \pasp, 122, 935, \dodoi{10.1086/655938}

\bibitem[{Foreman-Mackey {et~al.}(2020)Foreman-Mackey, Luger, Czekala, Agol, Price-Whelan, \& Barclay}]{exoplanet:exoplanet}
Foreman-Mackey, D., Luger, R., Czekala, I., {et~al.} 2020, exoplanet-dev/exoplanet v0.3.2, \dodoi{10.5281/zenodo.1998447}

\bibitem[{{Fulton} {et~al.}(2018){Fulton}, {Petigura}, {Blunt}, \& {Sinukoff}}]{radvel}
{Fulton}, B.~J., {Petigura}, E.~A., {Blunt}, S., \& {Sinukoff}, E. 2018, \pasp, 130, 044504, \dodoi{10.1088/1538-3873/aaaaa8}

\bibitem[{{Gaia Collaboration} {et~al.}(2016){Gaia Collaboration}, {Prusti}, {de Bruijne}, {Brown}, {Vallenari}, {Babusiaux}, {Bailer-Jones}, {Bastian}, {Biermann}, {Evans}, {Eyer}, {Jansen}, {Jordi}, {Klioner}, {Lammers}, {Lindegren}, {Luri}, {Mignard}, {Milligan}, {Panem}, {Poinsignon}, {Pourbaix}, {Randich}, {Sarri}, {Sartoretti}, {Siddiqui}, {Soubiran}, {Valette}, {van Leeuwen}, {Walton}, {Aerts}, {Arenou}, {Cropper}, {Drimmel}, {H{\o}g}, {Katz}, {Lattanzi}, {O'Mullane}, {Grebel}, {Holland}, {Huc}, {Passot}, {Bramante}, {Cacciari}, {Casta{\~n}eda}, {Chaoul}, {Cheek}, {De Angeli}, {Fabricius}, {Guerra}, {Hern{\'a}ndez}, {Jean-Antoine-Piccolo}, {Masana}, {Messineo}, {Mowlavi}, {Nienartowicz}, {Ord{\'o}{\~n}ez-Blanco}, {Panuzzo}, {Portell}, {Richards}, {Riello}, {Seabroke}, {Tanga}, {Th{\'e}venin}, {Torra}, {Els}, {Gracia-Abril}, {Comoretto}, {Garcia-Reinaldos}, {Lock}, {Mercier}, {Altmann}, {Andrae}, {Astraatmadja}, {Bellas-Velidis}, {Benson}, {Berthier}, {Blomme}, {Busso}, {Carry}, {Cellino}, {Clementini},
  {Cowell}, {Creevey}, {Cuypers}, {Davidson}, {De Ridder}, {de Torres}, {Delchambre}, {Dell'Oro}, {Ducourant}, {Fr{\'e}mat}, {Garc{\'\i}a-Torres}, {Gosset}, {Halbwachs}, {Hambly}, {Harrison}, {Hauser}, {Hestroffer}, {Hodgkin}, {Huckle}, {Hutton}, {Jasniewicz}, {Jordan}, {Kontizas}, {Korn}, {Lanzafame}, {Manteiga}, {Moitinho}, {Muinonen}, {Osinde}, {Pancino}, {Pauwels}, {Petit}, {Recio-Blanco}, {Robin}, {Sarro}, {Siopis}, {Smith}, {Smith}, {Sozzetti}, {Thuillot}, {van Reeven}, {Viala}, {Abbas}, {Abreu Aramburu}, {Accart}, {Aguado}, {Allan}, {Allasia}, {Altavilla}, {{\'A}lvarez}, {Alves}, {Anderson}, {Andrei}, {Anglada Varela}, {Antiche}, {Antoja}, {Ant{\'o}n}, {Arcay}, {Atzei}, {Ayache}, {Bach}, {Baker}, {Balaguer-N{\'u}{\~n}ez}, {Barache}, {Barata}, {Barbier}, {Barblan}, {Baroni}, {Barrado y Navascu{\'e}s}, {Barros}, {Barstow}, {Becciani}, {Bellazzini}, {Bellei}, {Bello Garc{\'\i}a}, {Belokurov}, {Bendjoya}, {Berihuete}, {Bianchi}, {Bienaym{\'e}}, {Billebaud}, {Blagorodnova}, {Blanco-Cuaresma}, {Boch},
  {Bombrun}, {Borrachero}, {Bouquillon}, {Bourda}, {Bouy}, {Bragaglia}, {Breddels}, {Brouillet}, {Br{\"u}semeister}, {Bucciarelli}, {Budnik}, {Burgess}, {Burgon}, {Burlacu}, {Busonero}, {Buzzi}, {Caffau}, {Cambras}, {Campbell}, {Cancelliere}, {Cantat-Gaudin}, {Carlucci}, {Carrasco}, {Castellani}, {Charlot}, {Charnas}, {Charvet}, {Chassat}, {Chiavassa}, {Clotet}, {Cocozza}, {Collins}, {Collins}, {Costigan}, {Crifo}, {Cross}, {Crosta}, {Crowley}, {Dafonte}, {Damerdji}, {Dapergolas}, {David}, {David}, {De Cat}, {de Felice}, {de Laverny}, {De Luise}, {De March}, {de Martino}, {de Souza}, {Debosscher}, {del Pozo}, {Delbo}, {Delgado}, {Delgado}, {di Marco}, {Di Matteo}, {Diakite}, {Distefano}, {Dolding}, {Dos Anjos}, {Drazinos}, {Dur{\'a}n}, {Dzigan}, {Ecale}, {Edvardsson}, {Enke}, {Erdmann}, {Escolar}, {Espina}, {Evans}, {Eynard Bontemps}, {Fabre}, {Fabrizio}, {Faigler}, {Falc{\~a}o}, {Farr{\`a}s Casas}, {Faye}, {Federici}, {Fedorets}, {Fern{\'a}ndez-Hern{\'a}ndez}, {Fernique}, {Fienga}, {Figueras}, {Filippi},
  {Findeisen}, {Fonti}, {Fouesneau}, {Fraile}, {Fraser}, {Fuchs}, {Furnell}, {Gai}, {Galleti}, {Galluccio}, {Garabato}, {Garc{\'\i}a-Sedano}, {Gar{\'e}}, {Garofalo}, {Garralda}, {Gavras}, {Gerssen}, {Geyer}, {Gilmore}, {Girona}, {Giuffrida}, {Gomes}, {Gonz{\'a}lez-Marcos}, {Gonz{\'a}lez-N{\'u}{\~n}ez}, {Gonz{\'a}lez-Vidal}, {Granvik}, {Guerrier}, {Guillout}, {Guiraud}, {G{\'u}rpide}, {Guti{\'e}rrez-S{\'a}nchez}, {Guy}, {Haigron}, {Hatzidimitriou}, {Haywood}, {Heiter}, {Helmi}, {Hobbs}, {Hofmann}, {Holl}, {Holland }, {Hunt}, {Hypki}, {Icardi}, {Irwin}, {Jevardat de Fombelle}, {Jofr{\'e}}, {Jonker}, {Jorissen}, {Julbe}, {Karampelas}, {Kochoska}, {Kohley}, {Kolenberg}, {Kontizas}, {Koposov}, {Kordopatis}, {Koubsky}, {Kowalczyk}, {Krone-Martins}, {Kudryashova}, {Kull}, {Bachchan}, {Lacoste-Seris}, {Lanza}, {Lavigne}, {Le Poncin-Lafitte}, {Lebreton}, {Lebzelter}, {Leccia}, {Leclerc}, {Lecoeur-Taibi}, {Lemaitre}, {Lenhardt}, {Leroux}, {Liao}, {Licata}, {Lindstr{\o}m}, {Lister}, {Livanou}, {Lobel}, {L{\"o}ffler},
  {L{\'o}pez}, {Lopez-Lozano}, {Lorenz}, {Loureiro}, {MacDonald}, {Magalh{\~a}es Fernandes}, {Managau}, {Mann}, {Mantelet}, {Marchal}, {Marchant}, {Marconi}, {Marie}, {Marinoni}, {Marrese}, {Marschalk{\'o}}, {Marshall}, {Mart{\'\i}n-Fleitas}, {Martino}, {Mary}, {Matijevi{\v{c}}}, {Mazeh}, {McMillan}, {Messina}, {Mestre}, {Michalik}, {Millar}, {Miranda}, {Molina}, {Molinaro}, {Molinaro}, {Moln{\'a}r}, {Moniez}, {Montegriffo}, {Monteiro}, {Mor}, {Mora}, {Morbidelli}, {Morel}, {Morgenthaler}, {Morley}, {Morris}, {Mulone}, {Muraveva}, {Musella}, {Narbonne}, {Nelemans}, {Nicastro}, {Noval}, {Ord{\'e}novic}, {Ordieres-Mer{\'e}}, {Osborne}, {Pagani}, {Pagano}, {Pailler}, {Palacin}, {Palaversa}, {Parsons}, {Paulsen}, {Pecoraro}, {Pedrosa}, {Pentik{\"a}inen}, {Pereira}, {Pichon}, {Piersimoni}, {Pineau}, {Plachy}, {Plum}, {Poujoulet}, {Pr{\v{s}}a}, {Pulone}, {Ragaini}, {Rago}, {Rambaux}, {Ramos-Lerate}, {Ranalli}, {Rauw}, {Read}, {Regibo}, {Renk}, {Reyl{\'e}}, {Ribeiro}, {Rimoldini}, {Ripepi}, {Riva}, {Rixon},
  {Roelens}, {Romero-G{\'o}mez}, {Rowell}, {Royer}, {Rudolph}, {Ruiz-Dern}, {Sadowski}, {Sagrist{\`a} Sell{\'e}s}, {Sahlmann}, {Salgado}, {Salguero}, {Sarasso}, {Savietto}, {Schnorhk}, {Schultheis}, {Sciacca}, {Segol}, {Segovia}, {Segransan}, {Serpell}, {Shih}, {Smareglia}, {Smart}, {Smith}, {Solano}, {Solitro}, {Sordo}, {Soria Nieto}, {Souchay}, {Spagna}, {Spoto}, {Stampa}, {Steele}, {Steidelm{\"u}ller}, {Stephenson}, {Stoev}, {Suess}, {S{\"u}veges}, {Surdej}, {Szabados}, {Szegedi-Elek}, {Tapiador}, {Taris}, {Tauran}, {Taylor}, {Teixeira}, {Terrett}, {Tingley}, {Trager}, {Turon}, {Ulla}, {Utrilla}, {Valentini}, {van Elteren}, {Van Hemelryck}, {van Leeuwen}, {Varadi}, {Vecchiato}, {Veljanoski}, {Via}, {Vicente}, {Vogt}, {Voss}, {Votruba}, {Voutsinas}, {Walmsley}, {Weiler}, {Weingrill}, {Werner}, {Wevers}, {Whitehead}, {Wyrzykowski}, {Yoldas}, {{\v{Z}}erjal}, {Zucker}, {Zurbach}, {Zwitter}, {Alecu}, {Allen}, {Allende Prieto}, {Amorim}, {Anglada-Escud{\'e}}, {Arsenijevic}, {Azaz}, {Balm}, {Beck}, {Bernstein},
  {Bigot}, {Bijaoui}, {Blasco}, {Bonfigli}, {Bono}, {Boudreault}, {Bressan}, {Brown}, {Brunet}, {Bunclark}, {Buonanno}, {Butkevich}, {Carret}, {Carrion}, {Chemin}, {Ch{\'e}reau}, {Corcione}, {Darmigny}, {de Boer}, {de Teodoro}, {de Zeeuw}, {Delle Luche}, {Domingues}, {Dubath}, {Fodor}, {Fr{\'e}zouls}, {Fries}, {Fustes}, {Fyfe}, {Gallardo}, {Gallegos}, {Gardiol}, {Gebran}, {Gomboc}, {G{\'o}mez}, {Grux}, {Gueguen}, {Heyrovsky}, {Hoar}, {Iannicola}, {Isasi Parache}, {Janotto}, {Joliet}, {Jonckheere}, {Keil}, {Kim}, {Klagyivik}, {Klar}, {Knude}, {Kochukhov}, {Kolka}, {Kos}, {Kutka}, {Lainey}, {LeBouquin}, {Liu}, {Loreggia}, {Makarov}, {Marseille}, {Martayan}, {Martinez-Rubi}, {Massart}, {Meynadier}, {Mignot}, {Munari}, {Nguyen}, {Nordlander}, {Ocvirk}, {O'Flaherty}, {Olias Sanz}, {Ortiz}, {Osorio}, {Oszkiewicz}, {Ouzounis}, {Palmer}, {Park}, {Pasquato}, {Peltzer}, {Peralta}, {P{\'e}turaud}, {Pieniluoma}, {Pigozzi}, {Poels}, {Prat}, {Prod'homme}, {Raison}, {Rebordao}, {Risquez}, {Rocca-Volmerange}, {Rosen},
  {Ruiz-Fuertes}, {Russo}, {Sembay}, {Serraller Vizcaino}, {Short}, {Siebert}, {Silva}, {Sinachopoulos}, {Slezak}, {Soffel}, {Sosnowska}, {Strai{\v{z}}ys}, {ter Linden}, {Terrell}, {Theil}, {Tiede}, {Troisi}, {Tsalmantza}, {Tur}, {Vaccari}, {Vachier}, {Valles}, {Van Hamme}, {Veltz}, {Virtanen}, {Wallut}, {Wichmann}, {Wilkinson}, {Ziaeepour}, \& {Zschocke}}]{gaia}
{Gaia Collaboration}, {Prusti}, T., {de Bruijne}, J.~H.~J., {et~al.} 2016, \aap, 595, A1, \dodoi{10.1051/0004-6361/201629272}

\bibitem[{{Gaia Collaboration} {et~al.}(2023){Gaia Collaboration}, {Vallenari}, {Brown}, {Prusti}, {de Bruijne}, {Arenou}, {Babusiaux}, {Biermann}, {Creevey}, {Ducourant}, {Evans}, {Eyer}, {Guerra}, {Hutton}, {Jordi}, {Klioner}, {Lammers}, {Lindegren}, {Luri}, {Mignard}, {Panem}, {Pourbaix}, {Randich}, {Sartoretti}, {Soubiran}, {Tanga}, {Walton}, {Bailer-Jones}, {Bastian}, {Drimmel}, {Jansen}, {Katz}, {Lattanzi}, {van Leeuwen}, {Bakker}, {Cacciari}, {Casta{\~n}eda}, {De Angeli}, {Fabricius}, {Fouesneau}, {Fr{\'e}mat}, {Galluccio}, {Guerrier}, {Heiter}, {Masana}, {Messineo}, {Mowlavi}, {Nicolas}, {Nienartowicz}, {Pailler}, {Panuzzo}, {Riclet}, {Roux}, {Seabroke}, {Sordo}, {Th{\'e}venin}, {Gracia-Abril}, {Portell}, {Teyssier}, {Altmann}, {Andrae}, {Audard}, {Bellas-Velidis}, {Benson}, {Berthier}, {Blomme}, {Burgess}, {Busonero}, {Busso}, {C{\'a}novas}, {Carry}, {Cellino}, {Cheek}, {Clementini}, {Damerdji}, {Davidson}, {de Teodoro}, {Nu{\~n}ez Campos}, {Delchambre}, {Dell'Oro}, {Esquej},
  {Fern{\'a}ndez-Hern{\'a}ndez}, {Fraile}, {Garabato}, {Garc{\'\i}a-Lario}, {Gosset}, {Haigron}, {Halbwachs}, {Hambly}, {Harrison}, {Hern{\'a}ndez}, {Hestroffer}, {Hodgkin}, {Holl}, {Jan{\ss}en}, {Jevardat de Fombelle}, {Jordan}, {Krone-Martins}, {Lanzafame}, {L{\"o}ffler}, {Marchal}, {Marrese}, {Moitinho}, {Muinonen}, {Osborne}, {Pancino}, {Pauwels}, {Recio-Blanco}, {Reyl{\'e}}, {Riello}, {Rimoldini}, {Roegiers}, {Rybizki}, {Sarro}, {Siopis}, {Smith}, {Sozzetti}, {Utrilla}, {van Leeuwen}, {Abbas}, {{\'A}brah{\'a}m}, {Abreu Aramburu}, {Aerts}, {Aguado}, {Ajaj}, {Aldea-Montero}, {Altavilla}, {{\'A}lvarez}, {Alves}, {Anders}, {Anderson}, {Anglada Varela}, {Antoja}, {Baines}, {Baker}, {Balaguer-N{\'u}{\~n}ez}, {Balbinot}, {Balog}, {Barache}, {Barbato}, {Barros}, {Barstow}, {Bartolom{\'e}}, {Bassilana}, {Bauchet}, {Becciani}, {Bellazzini}, {Berihuete}, {Bernet}, {Bertone}, {Bianchi}, {Binnenfeld}, {Blanco-Cuaresma}, {Blazere}, {Boch}, {Bombrun}, {Bossini}, {Bouquillon}, {Bragaglia}, {Bramante}, {Breedt},
  {Bressan}, {Brouillet}, {Brugaletta}, {Bucciarelli}, {Burlacu}, {Butkevich}, {Buzzi}, {Caffau}, {Cancelliere}, {Cantat-Gaudin}, {Carballo}, {Carlucci}, {Carnerero}, {Carrasco}, {Casamiquela}, {Castellani}, {Castro-Ginard}, {Chaoul}, {Charlot}, {Chemin}, {Chiaramida}, {Chiavassa}, {Chornay}, {Comoretto}, {Contursi}, {Cooper}, {Cornez}, {Cowell}, {Crifo}, {Cropper}, {Crosta}, {Crowley}, {Dafonte}, {Dapergolas}, {David}, {David}, {de Laverny}, {De Luise}, {De March}, {De Ridder}, {de Souza}, {de Torres}, {del Peloso}, {del Pozo}, {Delbo}, {Delgado}, {Delisle}, {Demouchy}, {Dharmawardena}, {Di Matteo}, {Diakite}, {Diener}, {Distefano}, {Dolding}, {Edvardsson}, {Enke}, {Fabre}, {Fabrizio}, {Faigler}, {Fedorets}, {Fernique}, {Fienga}, {Figueras}, {Fournier}, {Fouron}, {Fragkoudi}, {Gai}, {Garcia-Gutierrez}, {Garcia-Reinaldos}, {Garc{\'\i}a-Torres}, {Garofalo}, {Gavel}, {Gavras}, {Gerlach}, {Geyer}, {Giacobbe}, {Gilmore}, {Girona}, {Giuffrida}, {Gomel}, {Gomez}, {Gonz{\'a}lez-N{\'u}{\~n}ez},
  {Gonz{\'a}lez-Santamar{\'\i}a}, {Gonz{\'a}lez-Vidal}, {Granvik}, {Guillout}, {Guiraud}, {Guti{\'e}rrez-S{\'a}nchez}, {Guy}, {Hatzidimitriou}, {Hauser}, {Haywood}, {Helmer}, {Helmi}, {Sarmiento}, {Hidalgo}, {Hilger}, {H{\l}adczuk}, {Hobbs}, {Holland}, {Huckle}, {Jardine}, {Jasniewicz}, {Jean-Antoine Piccolo}, {Jim{\'e}nez-Arranz}, {Jorissen}, {Juaristi Campillo}, {Julbe}, {Karbevska}, {Kervella}, {Khanna}, {Kontizas}, {Kordopatis}, {Korn}, {K{\'o}sp{\'a}l}, {Kostrzewa-Rutkowska}, {Kruszy{\'n}ska}, {Kun}, {Laizeau}, {Lambert}, {Lanza}, {Lasne}, {Le Campion}, {Lebreton}, {Lebzelter}, {Leccia}, {Leclerc}, {Lecoeur-Taibi}, {Liao}, {Licata}, {Lindstr{\o}m}, {Lister}, {Livanou}, {Lobel}, {Lorca}, {Loup}, {Madrero Pardo}, {Magdaleno Romeo}, {Managau}, {Mann}, {Manteiga}, {Marchant}, {Marconi}, {Marcos}, {Marcos Santos}, {Mar{\'\i}n Pina}, {Marinoni}, {Marocco}, {Marshall}, {Martin Polo}, {Mart{\'\i}n-Fleitas}, {Marton}, {Mary}, {Masip}, {Massari}, {Mastrobuono-Battisti}, {Mazeh}, {McMillan}, {Messina}, {Michalik},
  {Millar}, {Mints}, {Molina}, {Molinaro}, {Moln{\'a}r}, {Monari}, {Mongui{\'o}}, {Montegriffo}, {Montero}, {Mor}, {Mora}, {Morbidelli}, {Morel}, {Morris}, {Muraveva}, {Murphy}, {Musella}, {Nagy}, {Noval}, {Oca{\~n}a}, {Ogden}, {Ordenovic}, {Osinde}, {Pagani}, {Pagano}, {Palaversa}, {Palicio}, {Pallas-Quintela}, {Panahi}, {Payne-Wardenaar}, {Pe{\~n}alosa Esteller}, {Penttil{\"a}}, {Pichon}, {Piersimoni}, {Pineau}, {Plachy}, {Plum}, {Poggio}, {Pr{\v{s}}a}, {Pulone}, {Racero}, {Ragaini}, {Rainer}, {Raiteri}, {Rambaux}, {Ramos}, {Ramos-Lerate}, {Re Fiorentin}, {Regibo}, {Richards}, {Rios Diaz}, {Ripepi}, {Riva}, {Rix}, {Rixon}, {Robichon}, {Robin}, {Robin}, {Roelens}, {Rogues}, {Rohrbasser}, {Romero-G{\'o}mez}, {Rowell}, {Royer}, {Ruz Mieres}, {Rybicki}, {Sadowski}, {S{\'a}ez N{\'u}{\~n}ez}, {Sagrist{\`a} Sell{\'e}s}, {Sahlmann}, {Salguero}, {Samaras}, {Sanchez Gimenez}, {Sanna}, {Santove{\~n}a}, {Sarasso}, {Schultheis}, {Sciacca}, {Segol}, {Segovia}, {S{\'e}gransan}, {Semeux}, {Shahaf}, {Siddiqui}, {Siebert},
  {Siltala}, {Silvelo}, {Slezak}, {Slezak}, {Smart}, {Snaith}, {Solano}, {Solitro}, {Souami}, {Souchay}, {Spagna}, {Spina}, {Spoto}, {Steele}, {Steidelm{\"u}ller}, {Stephenson}, {S{\"u}veges}, {Surdej}, {Szabados}, {Szegedi-Elek}, {Taris}, {Taylor}, {Teixeira}, {Tolomei}, {Tonello}, {Torra}, {Torra}, {Torralba Elipe}, {Trabucchi}, {Tsounis}, {Turon}, {Ulla}, {Unger}, {Vaillant}, {van Dillen}, {van Reeven}, {Vanel}, {Vecchiato}, {Viala}, {Vicente}, {Voutsinas}, {Weiler}, {Wevers}, {Wyrzykowski}, {Yoldas}, {Yvard}, {Zhao}, {Zorec}, {Zucker}, \& {Zwitter}}]{gaiadr3}
{Gaia Collaboration}, {Vallenari}, A., {Brown}, A.~G.~A., {et~al.} 2023, \aap, 674, A1, \dodoi{10.1051/0004-6361/202243940}

\bibitem[{{Gan} {et~al.}(2021){Gan}, {Wang}, {Teske}, {Mao}, {Howard}, {Law}, {Batalha}, {Vanderburg}, {Dragomir}, {Huang}, {Feng}, {Butler}, {Crane}, {Shectman}, {Beletsky}, {Shporer}, {Montet}, {Burt}, {Feinstein}, {Flowers}, {Nandakumar}, {Barbieri}, {Corbett}, {Ratzloff}, {Galliher}, {Chavez}, {Vasquez}, {Glazier}, \& {Haislip}}]{gan21}
{Gan}, T., {Wang}, S.~X., {Teske}, J.~K., {et~al.} 2021, \mnras, 501, 6042, \dodoi{10.1093/mnras/staa3886}

\bibitem[{{Gelman} \& {Rubin}(1992)}]{gelman92}
{Gelman}, A., \& {Rubin}, D.~B. 1992, Statistical Science, 7, 457, \dodoi{10.1214/ss/1177011136}

\bibitem[{{Grunblatt} {et~al.}(2015){Grunblatt}, {Howard}, \& {Haywood}}]{grunblatt15}
{Grunblatt}, S.~K., {Howard}, A.~W., \& {Haywood}, R.~D. 2015, \apj, 808, 127, \dodoi{10.1088/0004-637X/808/2/127}

\bibitem[{{Guerrero} {et~al.}(2021){Guerrero}, {Seager}, {Huang}, {Vanderburg}, {Garcia Soto}, {Mireles}, {Hesse}, {Fong}, {Glidden}, {Shporer}, {Latham}, {Collins}, {Quinn}, {Burt}, {Dragomir}, {Crossfield}, {Vanderspek}, {Fausnaugh}, {Burke}, {Ricker}, {Daylan}, {Essack}, {G{\"u}nther}, {Osborn}, {Pepper}, {Rowden}, {Sha}, {Villanueva}, {Yahalomi}, {Yu}, {Ballard}, {Batalha}, {Berardo}, {Chontos}, {Dittmann}, {Esquerdo}, {Mikal-Evans}, {Jayaraman}, {Krishnamurthy}, {Louie}, {Mehrle}, {Niraula}, {Rackham}, {Rodriguez}, {Rowden}, {Sousa-Silva}, {Watanabe}, {Wong}, {Zhan}, {Zivanovic}, {Christiansen}, {Ciardi}, {Swain}, {Lund}, {Mullally}, {Fleming}, {Rodriguez}, {Boyd}, {Quintana}, {Barclay}, {Col{\'o}n}, {Rinehart}, {Schlieder}, {Clampin}, {Jenkins}, {Twicken}, {Caldwell}, {Coughlin}, {Henze}, {Lissauer}, {Morris}, {Rose}, {Smith}, {Tenenbaum}, {Ting}, {Wohler}, {Bakos}, {Bean}, {Berta-Thompson}, {Bieryla}, {Bouma}, {Buchhave}, {Butler}, {Charbonneau}, {Doty}, {Ge}, {Holman}, {Howard}, {Kaltenegger}, {Kane},
  {Kjeldsen}, {Kreidberg}, {Lin}, {Minsky}, {Narita}, {Paegert}, {P{\'a}l}, {Palle}, {Sasselov}, {Spencer}, {Sozzetti}, {Stassun}, {Torres}, {Udry}, \& {Winn}}]{guerrero21}
{Guerrero}, N.~M., {Seager}, S., {Huang}, C.~X., {et~al.} 2021, \apjs, 254, 39, \dodoi{10.3847/1538-4365/abefe1}

\bibitem[{{Hadden} \& {Lithwick}(2014)}]{haddenLithwick14}
{Hadden}, S., \& {Lithwick}, Y. 2014, \apj, 787, 80, \dodoi{10.1088/0004-637X/787/1/80}

\bibitem[{Harris {et~al.}(2020)Harris, Millman, van~der Walt, Gommers, Virtanen, Cournapeau, Wieser, Taylor, Berg, Smith, Kern, Picus, Hoyer, van Kerkwijk, Brett, Haldane, del R{'{\i}}o, Wiebe, Peterson, G{'{e}}rard-Marchant, Sheppard, Reddy, Weckesser, Abbasi, Gohlke, \& Oliphant}]{numpy}
Harris, C.~R., Millman, K.~J., van~der Walt, S.~J., {et~al.} 2020, Nature, 585, 357, \dodoi{10.1038/s41586-020-2649-2}

\bibitem[{Hoffman \& Gelman(2014)}]{nuts:hoffman14}
Hoffman, M., \& Gelman, A. 2014, J. Mach. Learn. Res., 15, 1593

\bibitem[{{Horne} \& {Baliunas}(1986)}]{horne86}
{Horne}, J.~H., \& {Baliunas}, S.~L. 1986, \apj, 302, 757, \dodoi{10.1086/164037}

\bibitem[{{Houk} \& {Swift}(1999)}]{houk99}
{Houk}, N., \& {Swift}, C. 1999, Michigan Spectral Survey, 5, 0

\bibitem[{Howard \& Fulton(2016)}]{howard16}
Howard, A.~W., \& Fulton, B.~J. 2016, Publications of the Astronomical Society of the Pacific, 128, 114401, \dodoi{10.1088/1538-3873/128/969/114401}

\bibitem[{Howard {et~al.}(2010)Howard, Johnson, Marcy, Fischer, Wright, Bernat, Henry, Peek, Isaacson, Apps, Endl, Cochran, Valenti, Anderson, \& Piskunov}]{howard10}
Howard, A.~W., Johnson, J.~A., Marcy, G.~W., {et~al.} 2010, The Astrophysical Journal, 721, 1467, \dodoi{10.1088/0004-637x/721/2/1467}

\bibitem[{Howell {et~al.}(2014)Howell, Sobeck, Haas, Still, Barclay, Mullally, Troeltzsch, Aigrain, Bryson, Caldwell, {et~al.}}]{howell14}
Howell, S.~B., Sobeck, C., Haas, M., {et~al.} 2014, Publications of the Astronomical Society of the Pacific, 126, 398

\bibitem[{{Huber} {et~al.}(2016){Huber}, {Bryson}, {Haas}, {Barclay}, {Barentsen}, {Howell}, {Sharma}, {Stello}, \& {Thompson}}]{epic}
{Huber}, D., {Bryson}, S.~T., {Haas}, M.~R., {et~al.} 2016, \apjs, 224, 2, \dodoi{10.3847/0067-0049/224/1/2}

\bibitem[{Hunter(2007)}]{matplotlib}
Hunter, J.~D. 2007, Computing in Science \& Engineering, 9, 90, \dodoi{10.1109/MCSE.2007.55}

\bibitem[{{Isaacson} {et~al.}(2024){Isaacson}, {Howard}, {Fulton}, {Petigura}, {Weiss}, {Kane}, {Carter}, {Beard}, {Giacalone}, {Van Zandt}, {Murphy}, {Dai}, {Chontos}, {Polanski}, {Rice}, {Lubin}, {Brinkman}, {Rubenzahl}, {Blunt}, {Yee}, {MacDougall}, {Dalba}, {Tyler}, {Behmard}, {Angelo}, {Pidhorodetska}, {Mayo}, {Holcomb}, {Turtelboom}, {Hill}, {Bouma}, {Zhang}, {Crossfield}, \& {Saunders}}]{isaacson24}
{Isaacson}, H., {Howard}, A.~W., {Fulton}, B., {et~al.} 2024, \apjs, 274, 35, \dodoi{10.3847/1538-4365/ad676c}

\bibitem[{{Jontof-Hutter}(2019)}]{jontofhutter19}
{Jontof-Hutter}, D. 2019, Annual Review of Earth and Planetary Sciences, 47, 141, \dodoi{10.1146/annurev-earth-053018-060352}

\bibitem[{{Kempton} {et~al.}(2018){Kempton}, {Bean}, {Louie}, {Deming}, {Koll}, {Mansfield}, {Christiansen}, {L{\'o}pez-Morales}, {Swain}, {Zellem}, {Ballard}, {Barclay}, {Barstow}, {Batalha}, {Beatty}, {Berta-Thompson}, {Birkby}, {Buchhave}, {Charbonneau}, {Cowan}, {Crossfield}, {de Val-Borro}, {Doyon}, {Dragomir}, {Gaidos}, {Heng}, {Hu}, {Kane}, {Kreidberg}, {Mallonn}, {Morley}, {Narita}, {Nascimbeni}, {Pall{\'e}}, {Quintana}, {Rauscher}, {Seager}, {Shkolnik}, {Sing}, {Sozzetti}, {Stassun}, {Valenti}, \& {von Essen}}]{kempton18}
{Kempton}, E. M.~R., {Bean}, J.~L., {Louie}, D.~R., {et~al.} 2018, \pasp, 130, 114401, \dodoi{10.1088/1538-3873/aadf6f}

\bibitem[{{Kennedy} \& {Kenyon}(2008)}]{kennedy08}
{Kennedy}, G.~M., \& {Kenyon}, S.~J. 2008, \apj, 673, 502, \dodoi{10.1086/524130}

\bibitem[{{Kokubo} \& {Ida}(2002)}]{kokubo02}
{Kokubo}, E., \& {Ida}, S. 2002, \apj, 581, 666, \dodoi{10.1086/344105}

\bibitem[{{Kosiarek} {et~al.}(2021){Kosiarek}, {Berardo}, {Crossfield}, {Laguna}, {Piaulet}, {Akana Murphy}, {Howell}, {Henry}, {Isaacson}, {Fulton}, {Weiss}, {Petigura}, {Behmard}, {Hirsch}, {Teske}, {Burt}, {Mills}, {Chontos}, {Mo{\v{c}}nik}, {Howard}, {Werner}, {Livingston}, {Krick}, {Beichman}, {Gorjian}, {Kreidberg}, {Morley}, {Christiansen}, {Morales}, {Scott}, {Crane}, {Wang}, {Shectman}, {Rosenthal}, {Grunblatt}, {Rubenzahl}, {Dalba}, {Giacalone}, {Villanueva}, {Liu}, {Dai}, {Hill}, {Rice}, {Kane}, \& {Mayo}}]{kosiarek21}
{Kosiarek}, M.~R., {Berardo}, D.~A., {Crossfield}, I. J.~M., {et~al.} 2021, \aj, 161, 47, \dodoi{10.3847/1538-3881/abca39}

\bibitem[{{Kov{\'a}cs} {et~al.}(2002){Kov{\'a}cs}, {Zucker}, \& {Mazeh}}]{kovacs02}
{Kov{\'a}cs}, G., {Zucker}, S., \& {Mazeh}, T. 2002, \aap, 391, 369, \dodoi{10.1051/0004-6361:20020802}

\bibitem[{{Lacedelli} {et~al.}(2024){Lacedelli}, {Pall{\`e}}, {Luque}, {Cadieux}, {Akana Murphy}, {Murgas}, {Zapatero Osorio}, {Tabernero}, {Collins}, {Watkins}, {L'Heureux}, {Doyon}, {Jankowski}, {Nowak}, {Artigau}, {Batalha}, {Bean}, {Bouchy}, {Brady}, {Canto Martins}, {Carleo}, {Cointepas}, {Conti}, {Cook}, {Crossfield}, {Gonz{\`a}lez Hern{\`a}ndez}, {Lewin}, {Nari}, {Nielsen}, {Orell-Miquel}, {Parc}, {Schwarz}, {Srdoc}, \& {Van Eylen}}]{lacedelli24arxiv}
{Lacedelli}, G., {Pall{\`e}}, E., {Luque}, R., {et~al.} 2024, arXiv e-prints, arXiv:2409.11083, \dodoi{10.48550/arXiv.2409.11083}

\bibitem[{Lange {et~al.}(2024)Lange, Murphy, Batalha, Crossfield, Dressing, Fulton, Howard, Huber, Isaacson, Kane, Petigura, Robertson, Weiss, Behmard, Beard, Blunt, Brinkman, Chontos, Dai, Dalba, Fetherolf, Giacalone, Hill, Holcomb, Lubin, MacDougall, Mayo, MoÄnik, Pidhorodetska, Polanski, Rice, Rosenthal, Rubenzahl, Scarsdale, Turtelboom, Zandt, Ciardi, \& Boyle}]{lange24}
Lange, S., Murphy, J. M.~A., Batalha, N.~M., {et~al.} 2024, The Astronomical Journal, 167, 282, \dodoi{10.3847/1538-3881/ad34d9}

\bibitem[{{Lecavelier Des Etangs}(2007)}]{lecavelier07}
{Lecavelier Des Etangs}, A. 2007, \aap, 461, 1185, \dodoi{10.1051/0004-6361:20065014}

\bibitem[{{Lee} \& {Chiang}(2016)}]{lee16}
{Lee}, E.~J., \& {Chiang}, E. 2016, \apj, 817, 90, \dodoi{10.3847/0004-637X/817/2/90}

\bibitem[{{Libby-Roberts} {et~al.}(2020){Libby-Roberts}, {Berta-Thompson}, {D{\'e}sert}, {Masuda}, {Morley}, {Lopez}, {Deck}, {Fabrycky}, {Fortney}, {Line}, {Sanchis-Ojeda}, \& {Winn}}]{libbyroberts20}
{Libby-Roberts}, J.~E., {Berta-Thompson}, Z.~K., {D{\'e}sert}, J.-M., {et~al.} 2020, \aj, 159, 57, \dodoi{10.3847/1538-3881/ab5d36}

\bibitem[{{Lomb}(1976)}]{lomb76}
{Lomb}, N.~R. 1976, \apss, 39, 447, \dodoi{10.1007/BF00648343}

\bibitem[{{Lopez} \& {Fortney}(2014)}]{lopezforney14}
{Lopez}, E.~D., \& {Fortney}, J.~J. 2014, \apj, 792, 1, \dodoi{10.1088/0004-637X/792/1/1}

\bibitem[{{Luger} {et~al.}(2018){Luger}, {Kruse}, {Foreman-Mackey}, {Agol}, \& {Saunders}}]{everest:luger18}
{Luger}, R., {Kruse}, E., {Foreman-Mackey}, D., {Agol}, E., \& {Saunders}, N. 2018, \aj, 156, 99, \dodoi{10.3847/1538-3881/aad230}

\bibitem[{{Lundkvist} {et~al.}(2016){Lundkvist}, {Kjeldsen}, {Albrecht}, {Davies}, {Basu}, {Huber}, {Justesen}, {Karoff}, {Silva Aguirre}, {van Eylen}, {Vang}, {Arentoft}, {Barclay}, {Bedding}, {Campante}, {Chaplin}, {Christensen-Dalsgaard}, {Elsworth}, {Gilliland}, {Handberg}, {Hekker}, {Kawaler}, {Lund}, {Metcalfe}, {Miglio}, {Rowe}, {Stello}, {Tingley}, \& {White}}]{lundkvist16}
{Lundkvist}, M.~S., {Kjeldsen}, H., {Albrecht}, S., {et~al.} 2016, Nature Communications, 7, 11201, \dodoi{10.1038/ncomms11201}

\bibitem[{{Luque} \& {Pall{\'e}}(2022)}]{luque22}
{Luque}, R., \& {Pall{\'e}}, E. 2022, Science, 377, 1211, \dodoi{10.1126/science.abl7164}

\bibitem[{{Luque} {et~al.}(2019){Luque}, {Nowak}, {Pall{\'e}}, {Dai}, {Kaminski}, {Nagel}, {Hidalgo}, {Bauer}, {Lafarga}, {Livingston}, {Barrag{\'a}n}, {Hirano}, {Fridlund}, {Gandolfi}, {Justesen}, {Hjorth}, {Van Eylen}, {Winn}, {Esposito}, {Morales}, {Albrecht}, {Alonso}, {Amado}, {Beck}, {Caballero}, {Cabrera}, {Cochran}, {Csizmadia}, {Deeg}, {Eigm{\"u}ller}, {Endl}, {Erikson}, {Fukui}, {Grziwa}, {Guenther}, {Hatzes}, {Knudstrup}, {Korth}, {Lam}, {Lund}, {Mathur}, {Monta{\~n}es-Rodr{\'\i}guez}, {Narita}, {Nespral}, {Niraula}, {P{\"a}tzold}, {Persson}, {Prieto-Arranz}, {Quirrenbach}, {Rauer}, {Redfield}, {Reiners}, {Ribas}, \& {Smith}}]{luque19}
{Luque}, R., {Nowak}, G., {Pall{\'e}}, E., {et~al.} 2019, \aap, 623, A114, \dodoi{10.1051/0004-6361/201834952}

\bibitem[{{Mayor} {et~al.}(2003){Mayor}, {Pepe}, {Queloz}, {Bouchy}, {Rupprecht}, {Lo Curto}, {Avila}, {Benz}, {Bertaux}, {Bonfils}, {Dall}, {Dekker}, {Delabre}, {Eckert}, {Fleury}, {Gilliotte}, {Gojak}, {Guzman}, {Kohler}, {Lizon}, {Longinotti}, {Lovis}, {Megevand}, {Pasquini}, {Reyes}, {Sivan}, {Sosnowska}, {Soto}, {Udry}, {van Kesteren}, {Weber}, \& {Weilenmann}}]{mayor03}
{Mayor}, M., {Pepe}, F., {Queloz}, D., {et~al.} 2003, The Messenger, 114, 20

\bibitem[{{Morbidelli} {et~al.}(2015){Morbidelli}, {Lambrechts}, {Jacobson}, \& {Bitsch}}]{morbidelli15}
{Morbidelli}, A., {Lambrechts}, M., {Jacobson}, S., \& {Bitsch}, B. 2015, \icarus, 258, 418, \dodoi{10.1016/j.icarus.2015.06.003}

\bibitem[{{Munari} {et~al.}(2014){Munari}, {Henden}, {Frigo}, {Zwitter}, {Bienaym{\'e}}, {Bland-Hawthorn}, {Boeche}, {Freeman}, {Gibson}, {Gilmore}, {Grebel}, {Helmi}, {Kordopatis}, {Levine}, {Navarro}, {Parker}, {Reid}, {Seabroke}, {Siebert}, {Siviero}, {Smith}, {Steinmetz}, {Templeton}, {Terrell}, {Welch}, {Williams}, \& {Wyse}}]{munari14}
{Munari}, U., {Henden}, A., {Frigo}, A., {et~al.} 2014, \aj, 148, 81, \dodoi{10.1088/0004-6256/148/5/81}

\bibitem[{{Mustill} {et~al.}(2017){Mustill}, {Davies}, \& {Johansen}}]{mustill17}
{Mustill}, A.~J., {Davies}, M.~B., \& {Johansen}, A. 2017, \mnras, 468, 3000, \dodoi{10.1093/mnras/stx693}

\bibitem[{{NASA Exoplanet Archive}(2024)}]{nea}
{NASA Exoplanet Archive}. 2024, Planetary Systems, Version: 2024-09-02 17:46,  NExScI-Caltech/IPAC, \dodoi{10.26133/NEA12}

\bibitem[{Neal(2003)}]{neal03}
Neal, R.~M. 2003, The Annals of Statistics, 31, 705–767, \dodoi{10.1214/aos/1056562461}

\bibitem[{{Neal}(2012)}]{neal12}
{Neal}, R.~M. 2012, arXiv e-prints, arXiv:1206.1901.
\newblock \doarXiv{1206.1901}

\bibitem[{{Owen} \& {Wu}(2013)}]{owen13}
{Owen}, J.~E., \& {Wu}, Y. 2013, \apj, 775, 105, \dodoi{10.1088/0004-637X/775/2/105}

\bibitem[{pandas~development team(2020)}]{pandas}
pandas~development team, T. 2020, pandas-dev/pandas: Pandas, latest,  Zenodo, \dodoi{10.5281/zenodo.3509134}

\bibitem[{Petigura {et~al.}(2017)Petigura, Howard, Marcy, Johnson, Isaacson, Cargile, Hebb, Fulton, Weiss, Morton, Winn, Rogers, Sinukoff, Hirsch, \& Crossfield}]{specmatchsynth}
Petigura, E.~A., Howard, A.~W., Marcy, G.~W., {et~al.} 2017, The Astronomical Journal, 154, 107, \dodoi{10.3847/1538-3881/aa80de}

\bibitem[{{Quirrenbach} {et~al.}(2014){Quirrenbach}, {Amado}, {Caballero}, {Mundt}, {Reiners}, {Ribas}, {Seifert}, {Abril}, {Aceituno}, {Alonso-Floriano}, {Ammler-von Eiff}, {Antona Jim{\'e}nez}, {Anwand-Heerwart}, {Azzaro}, {Bauer}, {Barrado}, {Becerril}, {B{\'e}jar}, {Ben{\'\i}tez}, {Berdi{\~n}as}, {C{\'a}rdenas}, {Casal}, {Claret}, {Colom{\'e}}, {Cort{\'e}s-Contreras}, {Czesla}, {Doellinger}, {Dreizler}, {Feiz}, {Fern{\'a}ndez}, {Galad{\'\i}}, {G{\'a}lvez-Ortiz}, {Garc{\'\i}a-Piquer}, {Garc{\'\i}a-Vargas}, {Garrido}, {Gesa}, {G{\'o}mez Galera}, {Gonz{\'a}lez {\'A}lvarez}, {Gonz{\'a}lez Hern{\'a}ndez}, {Gr{\"o}zinger}, {Gu{\`a}rdia}, {Guenther}, {de Guindos}, {Guti{\'e}rrez-Soto}, {Hagen}, {Hatzes}, {Hauschildt}, {Helmling}, {Henning}, {Hermann}, {Hern{\'a}ndez Casta{\~n}o}, {Herrero}, {Hidalgo}, {Holgado}, {Huber}, {Huber}, {Jeffers}, {Joergens}, {de Juan}, {Kehr}, {Klein}, {K{\"u}rster}, {Lamert}, {Lalitha}, {Laun}, {Lemke}, {Lenzen}, {L{\'o}pez del Fresno}, {L{\'o}pez Mart{\'\i}}, {L{\'o}pez-Santiago},
  {Mall}, {Mandel}, {Mart{\'\i}n}, {Mart{\'\i}n-Ruiz}, {Mart{\'\i}nez-Rodr{\'\i}guez}, {Marvin}, {Mathar}, {Mirabet}, {Montes}, {Morales Mu{\~n}oz}, {Moya}, {Naranjo}, {Ofir}, {Oreiro}, {Pall{\'e}}, {Panduro}, {Passegger}, {P{\'e}rez-Calpena}, {P{\'e}rez Medialdea}, {Perger}, {Pluto}, {Ram{\'o}n}, {Rebolo}, {Redondo}, {Reffert}, {Reinhardt}, {Rhode}, {Rix}, {Rodler}, {Rodr{\'\i}guez}, {Rodr{\'\i}guez-L{\'o}pez}, {Rodr{\'\i}guez-P{\'e}rez}, {Rohloff}, {Rosich}, {S{\'a}nchez-Blanco}, {S{\'a}nchez Carrasco}, {Sanz-Forcada}, {Sarmiento}, {Sch{\"a}fer}, {Schiller}, {Schmidt}, {Schmitt}, {Solano}, {Stahl}, {Storz}, {St{\"u}rmer}, {Su{\'a}rez}, {Ulbrich}, {Veredas}, {Wagner}, {Winkler}, {Zapatero Osorio}, {Zechmeister}, {Abell{\'a}n de Paco}, {Anglada-Escud{\'e}}, {del Burgo}, {Klutsch}, {Lizon}, {L{\'o}pez-Morales}, {Morales}, {Perryman}, {Tulloch}, \& {Xu}}]{carmenes14}
{Quirrenbach}, A., {Amado}, P.~J., {Caballero}, J.~A., {et~al.} 2014, in Society of Photo-Optical Instrumentation Engineers (SPIE) Conference Series, Vol. 9147, Ground-based and Airborne Instrumentation for Astronomy V, ed. S.~K. {Ramsay}, I.~S. {McLean}, \& H.~{Takami}, 91471F, \dodoi{10.1117/12.2056453}

\bibitem[{{Quirrenbach} {et~al.}(2018){Quirrenbach}, {Amado}, {Ribas}, {Reiners}, {Caballero}, {Seifert}, {Aceituno}, {Azzaro}, {Baroch}, {Barrado}, {Bauer}, {Becerril}, {B{\`e}jar}, {Ben{\'\i}tez}, {Brinkm{\"o}ller}, {Cardona Guill{\'e}n}, {Cifuentes}, {Colom{\'e}}, {Cort{\'e}s-Contreras}, {Czesla}, {Dreizler}, {Fr{\"o}lich}, {Fuhrmeister}, {Galad{\'\i}-Enr{\'\i}quez}, {Gonz{\'a}lez Hern{\'a}ndez}, {Gonz{\'a}lez Peinado}, {Guenther}, {de Guindos}, {Hagen}, {Hatzes}, {Hauschildt}, {Helmling}, {Henning}, {Herbort}, {Hern{\'a}ndez Casta{\~n}o}, {Herrero}, {Hintz}, {Jeffers}, {Johnson}, {de Juan}, {Kaminski}, {Klahr}, {K{\"u}rster}, {Lafarga}, {Sairam}, {Lamp{\'o}n}, {Lara}, {Launhardt}, {L{\'o}pez del Fresno}, {L{\'o}pez-Puertas}, {Luque}, {Mandel}, {Marfil}, {Mart{\'\i}n}, {Mart{\'\i}n-Ruiz}, {Mathar}, {Montes}, {Morales}, {Nagel}, {Nortmann}, {Nowak}, {Pall{\'e}}, {Passegger}, {Pavlov}, {Pedraz}, {P{\'e}rez-Medialdea}, {Perger}, {Rebolo}, {Reffert}, {Rodr{\'\i}guez}, {Rodr{\'\i}guez L{\'o}pez}, {Rosich},
  {Sabotta}, {Sadegi}, {Salz}, {S{\'a}nchez-L{\'o}pez}, {Sanz-Forcada}, {Sarkis}, {Sch{\"a}fer}, {Schiller}, {Schmitt}, {Sch{\"o}fer}, {Schweitzer}, {Shulyak}, {Solano}, {Stahl}, {Tala Pinto}, {Trifonov}, {Zapatero Osorio}, {Yan}, {Zechmeister}, {Abell{\'a}n}, {Abril}, {Alonso-Floriano}, {Ammler-von Eiff}, {Anglada-Escud{\'e}}, {Anwand-Heerwart}, {Arroyo-Torres}, {Berdi{\~n}as}, {Bergondy}, {Bl{\"u}mcke}, {del Burgo}, {Cano}, {Carro}, {C{\'a}rdenas}, {Casal}, {Claret}, {D{\'\i}ez-Alonso}, {Doellinger}, {Dorda}, {Feiz}, {Fern{\'a}ndez}, {Ferro}, {Gaisn{\'e}}, {Gallardo}, {G{\'a}lvez-Ortiz}, {Garc{\'\i}a-Piquer}, {Garc{\'\i}a-Vargas}, {Garrido}, {Gesa}, {G{\'o}mez Galera}, {Gonz{\'a}lez-{\'A}lvarez}, {Gonz{\'a}lez-Cuesta}, {Grohnert}, {Gr{\"o}zinger}, {Gu{\`a}rdia}, {Guijarro}, {Hedrosa}, {Hermann}, {Hermelo}, {Hern{\'a}ndez Arab{\'\i}}, {Hern{\'a}ndez Hernando}, {Hidalgo}, {Holgado}, {Huber}, {Huber}, {Huke}, {Kehr}, {Kim}, {Klein}, {Kl{\"u}ter}, {Klutsch}, {Labarga}, {Labiche}, {Lamert}, {Laun}, {L{\'a}zaro},
  {Lemke}, {Lenzen}, {Llamas}, {Lizon}, {Lodieu}, {L{\'o}pez Gonz{\'a}lez}, {L{\'o}pez-Morales}, {L{\'o}pez Salas}, {L{\'o}pez-Santiago}, {Mag{\'a}n Madinabeitia}, {Mall}, {Mancini}, {Mar{\'\i}n Molina}, {Mart{\'\i}nez-Rodr{\'\i}guez}, {Maroto Fern{\'a}ndez}, {Marvin}, {Mirabet}, {Moreno-Raya}, {Moya}, {Mundt}, {Naranjo}, {Panduro}, {Pascual}, {P{\'e}rez-Calpena}, {Perryman}, {Pluto}, {Ram{\'o}n}, {Redondo}, {Reinhart}, {Rhode}, {Rix}, {Rodler}, {Rohloff}, {S{\'a}nchez-Blanco}, {S{\'a}nchez Carrasco}, {Sarmiento}, {Schmidt}, {Storz}, {Strachan}, {St{\"u}rmer}, {Su{\'a}rez}, {Tabernero}, {Tal-Or}, {Tulloch}, {Ulbrich}, {Veredas}, {Vico Linares}, {Vidal-Dasilva}, {Vilardell}, {Wagner}, {Winkler}, {Wolthoff}, {Xu}, \& {Zhao}}]{carmenes18}
{Quirrenbach}, A., {Amado}, P.~J., {Ribas}, I., {et~al.} 2018, in Society of Photo-Optical Instrumentation Engineers (SPIE) Conference Series, Vol. 10702, Ground-based and Airborne Instrumentation for Astronomy VII, ed. C.~J. {Evans}, L.~{Simard}, \& H.~{Takami}, 107020W, \dodoi{10.1117/12.2313689}

\bibitem[{{Rajpaul} {et~al.}(2017){Rajpaul}, {Buchhave}, \& {Aigrain}}]{rajpaul17}
{Rajpaul}, V., {Buchhave}, L.~A., \& {Aigrain}, S. 2017, \mnras, 471, L125, \dodoi{10.1093/mnrasl/slx116}

\bibitem[{{Rasmussen} \& {Williams}(2006)}]{rasmussen06}
{Rasmussen}, C.~E., \& {Williams}, C. K.~I. 2006, {Gaussian Processes for Machine Learning} (MIT Press)

\bibitem[{{Raymond} {et~al.}(2018){Raymond}, {Boulet}, {Izidoro}, {Esteves}, \& {Bitsch}}]{raymond18}
{Raymond}, S.~N., {Boulet}, T., {Izidoro}, A., {Esteves}, L., \& {Bitsch}, B. 2018, \mnras, 479, L81, \dodoi{10.1093/mnrasl/sly100}

\bibitem[{Ricker {et~al.}(2014)Ricker, Winn, Vanderspek, Latham, Bakos, Bean, Berta-Thompson, Brown, Buchhave, Butler, Butler, Chaplin, Charbonneau, Christensen-Dalsgaard, Clampin, Deming, Doty, Lee, Dressing, Dunham, Endl, Fressin, Ge, Henning, Holman, Howard, Ida, Jenkins, Jernigan, Johnson, Kaltenegger, Kawai, Kjeldsen, Laughlin, Levine, Lin, Lissauer, MacQueen, Marcy, McCullough, Morton, Narita, Paegert, Palle, Pepe, Pepper, Quirrenbach, Rinehart, Sasselov, Sato, Seager, Sozzetti, Stassun, Sullivan, Szentgyorgyi, Torres, Udry, \& Villasenor}]{ricker14}
Ricker, G.~R., Winn, J.~N., Vanderspek, R., {et~al.} 2014, Journal of Astronomical Telescopes, Instruments, and Systems, 1, 1 , \dodoi{10.1117/1.JATIS.1.1.014003}

\bibitem[{{Rosenthal} {et~al.}(2021){Rosenthal}, {Fulton}, {Hirsch}, {Isaacson}, {Howard}, {Dedrick}, {Sherstyuk}, {Blunt}, {Petigura}, {Knutson}, {Behmard}, {Chontos}, {Crepp}, {Crossfield}, {Dalba}, {Fischer}, {Henry}, {Kane}, {Kosiarek}, {Marcy}, {Rubenzahl}, {Weiss}, \& {Wright}}]{rosenthal21}
{Rosenthal}, L.~J., {Fulton}, B.~J., {Hirsch}, L.~A., {et~al.} 2021, \apjs, 255, 8, \dodoi{10.3847/1538-4365/abe23c}

\bibitem[{Salvatier {et~al.}(2016)Salvatier, Wiecki, \& Fonnesbeck}]{pymc3}
Salvatier, J., Wiecki, T.~V., \& Fonnesbeck, C. 2016, PeerJ Computer Science, 2, \dodoi{10.7717/peerj-cs.55}

\bibitem[{{Savitzky} \& {Golay}(1964)}]{savitzky64}
{Savitzky}, A., \& {Golay}, M.~J.~E. 1964, Analytical Chemistry, 36, 1627

\bibitem[{{Scargle}(1982)}]{scargle82}
{Scargle}, J.~D. 1982, \apj, 263, 835, \dodoi{10.1086/160554}

\bibitem[{{Schlichting}(2014)}]{schlichting14}
{Schlichting}, H.~E. 2014, \apjl, 795, L15, \dodoi{10.1088/2041-8205/795/1/L15}

\bibitem[{{Schwarz}(1978)}]{schwarz78}
{Schwarz}, G. 1978, Annals of Statistics, 6, 461

\bibitem[{{Sinukoff} {et~al.}(2017){Sinukoff}, {Howard}, {Petigura}, {Fulton}, {Crossfield}, {Isaacson}, {Gonzales}, {Crepp}, {Brewer}, {Hirsch}, {Weiss}, {Ciardi}, {Schlieder}, {Benneke}, {Christiansen}, {Dressing}, {Hansen}, {Knutson}, {Kosiarek}, {Livingston}, {Greene}, {Rogers}, \& {L{\'e}pine}}]{sinukoff17}
{Sinukoff}, E., {Howard}, A.~W., {Petigura}, E.~A., {et~al.} 2017, \aj, 153, 271, \dodoi{10.3847/1538-3881/aa725f}

\bibitem[{{Skrutskie} {et~al.}(2006){Skrutskie}, {Cutri}, {Stiening}, {Weinberg}, {Schneider}, {Carpenter}, {Beichman}, {Capps}, {Chester}, {Elias}, {Huchra}, {Liebert}, {Lonsdale}, {Monet}, {Price}, {Seitzer}, {Jarrett}, {Kirkpatrick}, {Gizis}, {Howard}, {Evans}, {Fowler}, {Fullmer}, {Hurt}, {Light}, {Kopan}, {Marsh}, {McCallon}, {Tam}, {Van Dyk}, \& {Wheelock}}]{2mass}
{Skrutskie}, M.~F., {Cutri}, R.~M., {Stiening}, R., {et~al.} 2006, \aj, 131, 1163, \dodoi{10.1086/498708}

\bibitem[{{Southworth}(2011)}]{southworth11}
{Southworth}, J. 2011, \mnras, 417, 2166, \dodoi{10.1111/j.1365-2966.2011.19399.x}

\bibitem[{{Szab{\'o}} \& {Kiss}(2011)}]{szabo11}
{Szab{\'o}}, G.~M., \& {Kiss}, L.~L. 2011, \apjl, 727, L44, \dodoi{10.1088/2041-8205/727/2/L44}

\bibitem[{Van~Rossum \& Drake(2009)}]{python3}
Van~Rossum, G., \& Drake, F.~L. 2009, Python 3 Reference Manual (Scotts Valley, CA: CreateSpace)

\bibitem[{{Vanderburg} {et~al.}(2016){Vanderburg}, {Plavchan}, {Johnson}, {Ciardi}, {Swift}, \& {Kane}}]{vanderburg16}
{Vanderburg}, A., {Plavchan}, P., {Johnson}, J.~A., {et~al.} 2016, \mnras, 459, 3565, \dodoi{10.1093/mnras/stw863}

\bibitem[{{Vehtari} {et~al.}(2021){Vehtari}, {Gelman}, {Simpson}, {Carpenter}, \& {B{\"u}rkner}}]{vehtari21}
{Vehtari}, A., {Gelman}, A., {Simpson}, D., {Carpenter}, B., \& {B{\"u}rkner}, P.-C. 2021, Bayesian Analysis, 16, 667, \dodoi{10.1214/20-BA1221}

\bibitem[{{Vogt} {et~al.}(1994){Vogt}, {Allen}, {Bigelow}, {Bresee}, {Brown}, {Cantrall}, {Conrad}, {Couture}, {Delaney}, {Epps}, {Hilyard}, {Hilyard}, {Horn}, {Jern}, {Kanto}, {Keane}, {Kibrick}, {Lewis}, {Osborne}, {Pardeilhan}, {Pfister}, {Ricketts}, {Robinson}, {Stover}, {Tucker}, {Ward}, \& {Wei}}]{vogt94}
{Vogt}, S.~S., {Allen}, S.~L., {Bigelow}, B.~C., {et~al.} 1994, in Society of Photo-Optical Instrumentation Engineers (SPIE) Conference Series, Vol. 2198, Instrumentation in Astronomy VIII, ed. D.~L. {Crawford} \& E.~R. {Craine}, 362, \dodoi{10.1117/12.176725}

\bibitem[{{Zechmeister} \& {K{\"u}rster}(2009)}]{zechmeister09}
{Zechmeister}, M., \& {K{\"u}rster}, M. 2009, \aap, 496, 577, \dodoi{10.1051/0004-6361:200811296}

\bibitem[{{Zeng} {et~al.}(2016){Zeng}, {Sasselov}, \& {Jacobsen}}]{zeng16}
{Zeng}, L., {Sasselov}, D.~D., \& {Jacobsen}, S.~B. 2016, \apj, 819, 127, \dodoi{10.3847/0004-637X/819/2/127}

\end{thebibliography}
\bibliographystyle{aasjournal}

\end{document}